\documentclass[12pt]{article}
\usepackage{lineno}
\usepackage[colorlinks, linkcolor=black, citecolor=black]{hyperref}
\usepackage{epsfig,epsf,psfrag,multirow,rotating}
\usepackage[latin1]{inputenc}
\usepackage{enumerate,enumitem,tikz}
\usepackage{lscape}
\usepackage{changepage,bbm,comment}
\usepackage{color,latexsym,graphicx}
\usepackage{amsfonts,amssymb,amscd,amsmath,amsbsy}
\usepackage{xcolor}
\usepackage{natbib}

\usepackage{mathtools}

\setlist[enumerate]{
  labelindent=0pt,
  leftmargin=*,
}

\setlist[itemize]{
  labelindent=0pt,
  leftmargin=*,
}

\usepackage{caption}
\usepackage{setspace} 

\usepackage{sectsty}
\sectionfont{\fontsize{12}{15}\selectfont}
\subsectionfont{\fontsize{12}{15}\selectfont}

\usepackage{amsmath}
\usepackage{times}
\usepackage{bm}
\usepackage{epsfig}
\usepackage{epsf}
\usepackage{psfrag}
\usepackage{natbib}
\usepackage{amssymb}
\usepackage{amsbsy}
\usepackage{lscape}
\usepackage{color}
\usepackage{latexsym}
\usepackage{amsfonts}
\usepackage{amscd}
\usepackage{graphicx}
\usepackage[latin1]{inputenc}
\usepackage{epstopdf}
\usepackage{mathrsfs}
\usepackage{rotating}
\usepackage{url}
\newcommand{\ba}{\begin{eqnarray}}
\newcommand{\ea}{\end{eqnarray}}
\newcommand{\bas}{\begin{eqnarray*}}
\newcommand{\eas}{\end{eqnarray*}}
\newcommand{\bit}{\begin{itemize}}
\newcommand{\eit}{\end{itemize}}
\newcommand{\ben}{\begin{enumerate}}
\newcommand{\een}{\end{enumerate}}

\newcommand{\e}{ { \mathbb{E}}}

\def\T{{ \mathrm{\scriptscriptstyle \top} }}

\newcommand{\pr}{ {\rm pr}}
\newcommand{\bone}{  \mbox{\bf 1}}
\newcommand{\bx}{  \mbox{\bf  x}}
\newcommand{\by}{  \mbox{\bf y}}
\newcommand{\bv}{  \mbox{\bf v}}
\newcommand{\bz}{  \mbox{\bf z}}

\newcommand{\bS}{  \mbox{\bf S}}
\newcommand{\bH}{  \mbox{\bf H}}
\newcommand{\bX}{  \mbox{\bf X}}
\newcommand{\bY}{  \mbox{\bf Y}}

\newcommand{\bbeta}{ \boldsymbol{\beta}}
\newcommand{\bvarphi}{ \boldsymbol{\varphi}}
\newcommand{\bgamma}{ \boldsymbol{\gamma}}
\newcommand{\bxi}{ \boldsymbol{\xi}}
\newcommand{\bSigma}{ \boldsymbol{\Sigma}}

\newcommand{\diag}{ {\rm diag}}
\newcommand{\pros}{  {\scriptsize\rm pros}}
\newcommand{\retr}{  {\scriptsize\rm retr}}

\newcommand{\sumsc}{ {\mbox{SS}} }
\newcommand{\fixedsc}{ \mbox{FS}}
\newcommand{\randomsc}{ \mbox{RS}}

\newtheorem{theorem}{Theorem}

\newtheorem{example}{Example}

\def\no{\noindent}

\definecolor{darkblue}{rgb}{0.0, 0.0, 0.5}

\mathtoolsset{showonlyrefs=true}

\begin{document}

{\centering {\large {\bf Retrospective score tests versus prospective score tests
for genetic association with case-control data}} \par}

\bigskip

\centerline{
Yukun Liu,
Pengfei Li,
Lei Song,
Kai Yu,
and Jing Qin 
}

\bigskip

\bigskip

\hrule

{\small
\begin{quotation}
\no
Since the seminal work by \cite{Prentice1979}, the prospective logistic likelihood
has become the standard method of analysis for retrospectively collected case-control data,
in particular for  testing the association between
a single genetic marker and a disease outcome in genetic case-control studies.
When studying multiple genetic markers with relatively small effects,
especially those with rare variants, various aggregated approaches
based on the same prospective likelihood have been developed to
integrate subtle association evidence among all considered markers.
In this paper we show that using the score statistic derived from
a prospective likelihood is not optimal in the analysis of retrospectively
sampled genetic data. We develop the locally most powerful genetic aggregation
test derived through the retrospective likelihood under a random effect model
assumption. In contrast to the fact that the disease prevalence information
 cannot be used to improve the efficiency for the estimation of odds ratio
 parameters in logistic regression models, we show that it can be
 utilized to enhance the testing power in genetic association studies.
 Extensive simulations demonstrate the advantages of the proposed method
 over the existing ones. One real genome-wide association study is analyzed for illustration.

\vspace{0.3cm}

\no
\end{quotation}
}

\hrule

\bigskip

\bigskip

\section{Introduction}
\label{introduction}

In  genetic association study of relatively rare disease outcome, such as rare cancer,
the case and control study is  probably one of the most commonly used designs  due  to
  its  convenience and cost  effectiveness.
In a case-control   design,  fixed  numbers of cases and controls are
ascertained for the covariate information. Given this information,
the most popular model for the disease status is the logistic regression model.
Since the seminal paper by \cite{Prentice1979},
it is  well known that one may use the prospective logistic likelihood to make
inference for the underlying odds ratio parameters
even  if data are collected retrospectively. In general the disease
prevalence is not estimable based on case and control data.
Since the intercept in the logistic regression and the disease
 prevalence are tangled together,
even if the knowledge of disease prevalence is available it can not be used to improve
the estimation of  the odds ratio parameters.
Many existing statistical genetic papers derived testing statistics
based on prospective logistic likelihood and
then applied them  without any justification  to the case and control data.
This strategy works most of times. In this paper, however,
we shall show that  in some applications
the score statistic derived from the retrospective likelihood
has a better power than that derived from prospective likelihood.

Denote the disease status as $D=0$ (non-disease) or $1$ (disease).
Let $X$ be a $d$-variate vector of  clinical covariates  and $Y$ be a
  $q$-variate vector representing measures on a set of considered genetic markers,
such as correlated markers within a candidate region or a gene.
A logistic regression model for the disease status given covariate information $(X=\bx, Y=\by)$ is
\[
\pr(D=1| \bx,  \by)=\pi(\alpha_p + \bx^\T \bbeta+ \by^\T\bgamma),
\]
where $\pi(t) = e^t/(1+e^t)$ is the logistic link function. The score test with respect to $\bgamma $
 can be used to  test the non-existence of genetic effect.
When the number of genetic markers  is relatively large, however,
the score test may loss power due to the increase in the degree-of-freedom.
To increase power,
one may reduce the dimension of $\bgamma$ based on certain assumption.
One commonly-used strategy is to assume   $\bgamma = \xi \bgamma_0$
where $\bgamma_0$ is a pre-specified vector and $\xi$ is the common genetic effect.
The simplest aggregation tests are the burden tests
(with, for example, $\bgamma_0=\bone$) and the adaptive burden tests, which collapse
information for multiple genetic variants into a single genetic score.
Clearly,  appropriately specified  $\bgamma_0$  could lead to more efficient test statistic.
In applications, $\bgamma_0$ is in general unknown.
  \cite{Lin2011} recommended to choose $K>1$ values of
$\bgamma_0$ and use the maximum of $K$ score statistics as their  test statistic.

The linear function  $\by^\T\bgamma$ of $\by$ in the  model
may not be general enough to capture more realistic scenarios,
where different genetic markers convey non-uniform risk levels (magnitude and or direction).
A popular approach is to utilize a random effect model
 \ba
 \label{random-effect-model}
 \pr(D=1|\bx, \by, v)=\pi ( \alpha_p + \bx^\T \bbeta  +  \by^\T
  \bgamma +  \by^\T  \bv \cdot \sqrt{\theta}  ),
\ea
where $ \bgamma$ and  $\bv$ denote respectively the fixed and random variant effects.
 Random effects are represented by $\sqrt{\theta}\bv$,
with the components of $\bv$ being independent and identically distributed from a distribution
with mean 0 and  a known variance.
If  $h(\bv)$ denotes the density function of $\bv$, we have a marginal probability
\bas
 \pr(D=1|\bx,  \by)=\int \pi\left( \alpha_p +  \bx^\T \bbeta
    +  \by^\T   \bgamma   + \by^\T  \bv \cdot \sqrt{\theta}  \right)h(\bv)d\bv,
\eas
which is no longer a logistic regression model any more.
If one specifies the random effect density $h(\bv)$, then it is possible
to derive the likelihood ratio statistic by numerical integration.
However the numerical integration is a formidable task,
especially when we examine thousands of genes  or regions.
The resulting likelihood ratio test would lose power when the random effect density is misspecified.

When assuming $\bgamma = 0$,  testing the non-existence of genetic effect
is equivalent to testing $H_0:\theta=0$.
The score test is very popular in statistical and genetic literature
since it is evaluated at the null $H_0:\theta=0$
and
can effectively avoid specifying the form of $h(\bv)$.
  \cite{Wu2011} proposed a sequence kernel association
test (SKAT), which   is a score test \citep{Lin1997}
under the assumption that all individuals
 in a study share a common but unobserved $\bv$.
Later on,  \cite{Lee2012b} extended  SKAT to SKAT-O,
 which  allows the components of $\bv$ to be equally correlated.
The burden, SKAT, and SKAT-O tests  focus either on the fixed effects or
on the random effects, and  hence may lose power when both the fixed and random effects
contribute to the overall genetic effects.
To enhance robustness and power,
\cite{Sun2013} proposed a mixed effects  score test (MiST) by
combining information from both fixed and random variant effects.

The SKAT, SKAT-O and MiST tests  are all built on
the assumption that observations from the individuals
share the same  random effect $\bv$
and are independent given $\bv$.
In other words, all subjects are not independent.
This assumption is fundamentally different
from the conventional random effect model assumption
(called the independent random effect assumption hereafter),
which treat observations as independent from each other unless
they come from the same individual or cluster
\citep{Verbeke1996,Wang1998,Ke2001,Jiang2007}.
This motivates us to develop score tests for the genetic effect based on case-control data
using the independent random effect model.
Specifically suppose that  $\{ (\bx_i, \by_i, D_i,  \bv_i): i=1, 2, \ldots, n\}$
are independent and identically distributed from model \eqref{random-effect-model}.
We observe not the random effects  $\bv_i$'s but   $\{ (\bx_i, \by_i, D_i): i=1, 2, \ldots, n\}$.
The  observations  are independent as the random effect is not shared among them.
For the convenience of presentation, we assume
$\bgamma = \gamma \bone$ and the components of $v_i$'s are uncorrelated with mean zero and variance 2.

The case-control study selects existing patients and then augments a certain number
of controls. Hence the numbers of patients and controls
are fixed. In contrast, those in a prospective study are random.
The proportion of patients  in a case-control study
may be severely different from that in the prospective sampling study.
For example, if the true disease prevalence is $5\%$, but one selects
2000 cases and 2000 controls in a case and control study,
then the proportions of proportions are $50\%$ and  $5\%$,
respectively for the case-control design and the prospective design.
This difference would lead to power loss for testing non-random effect
greatly if one does not distinguish these two different study designs.
To the best of our knowledge,  this phenomenon has never been observed
in  statistical or genetic literature.

 In this paper we will systematically investigate different score test statistics
 based on prospectively and retrospectively collected data under the independent random effect model.
 Connections and differences for testing   the non-existence of random effect
 between the case-control design and prospective design  will be highlighted.
 Interestingly we have found the information on disease  prevalence can be utilized
 to enhance the power for testing the existence of random effect.
 On the other hand if one blindly uses the score statistic derived from
 prospective likelihood for the retrospectively collected case and control data,
 then the test statistic may  lose power substantially.
 Methods developed in this paper may shed lights on other medical
 or econometric problems based on case and control data or
 choice-biased sampling data \citep{Amemiya1985}.

The rest of this paper is organized as follows.
Section 2 derives retrospective and prospective score tests with
respect to both fixed  and random effects based on case-control data.
The connection and difference between
the prospective and retrospective scores are made clear explicitly.
Section 3 establishes the asymptotical normalities of
all score statistics  derived in Section 2.
Based on the retrospective scores and their asymptotic normalities,
we construct several tests for   the overall
genetic effect   in Section 4.
A simulation study and a real data analysis are reported in Sections 5 and 6, respectively.
For convenience of presentation, we defer the technical details to the supplementary material.

\section{Retrospective and prospective score statistics}
\label{retro-prop}

Since case-control data are retrospectively collected data,
we first present the retrospective likelihood and
derive the retrospective score statistics for both the random- and fixed-effects.
Then we take them as if they were prospective data
and derive the corresponding prospective score statistics.

\subsection{Retrospective likelihood}
\label{retro-likelihood}

Let  $\bvarphi=(\alpha_p, \bbeta, \gamma, \theta)^\T$,
$f(\bx, \by)$ be the joint density function of $(\bx, \by)$, and
\bas
g(\bx, \by, \bvarphi)
=
\pr(D=1|\bx, \by) =
\int \pi( \alpha_p +   \bx^\T \bbeta  +  \by^\T \bone \gamma   + \by^\T  \bv \cdot \sqrt{\theta}  )  h(\bv)d\bv,
\eas
where
$h(\bv)$ is the density function of $\bv$. Using Bayes's formula,
 we can  find the densities of covariates in cases and controls are, respectively,
\bas
f_1(\bx, \by, \bvarphi)
 &:=& \pr(\bx, \by|D=1)
 = \frac{g(\bx, \by, \bvarphi)f(\bx, \by)}{\int g(\bx, \by, \bvarphi)f(\bx, \by)d\bx d\by},  \\
f_0(\bx, \by, \bvarphi)
 &:=& \pr(\bx, \by|D=0)
 = \frac{\{1-g(\bx, \by, \bvarphi)\}f(\bx, \by)}{1-  \int g(\bx, \by, \bvarphi)f(\bx, \by)d\bx d\by}.
\eas
Without loss of generality, we assume the first $n_0$  of  $\{ (\bx_i, \by_i, D_i): i=1, 2, \ldots, n\}$
are controls  and the last $n_1=n-n_0$ are cases.
The retrospective likelihood based on  case and control data is
\ba
\label{retro-likelihood}
L_{\retr}(\bvarphi)
&=& \prod_{i=1}^n  [   \{f_1(\bx_i, \by_i, \bvarphi)\}^{D_i} \{f_0(\bx_i, \by_i, \bvarphi)\}^{1-D_i}  ]
\ea
and the corresponding log likelihood is
\bas
\ell_{\retr}(\bvarphi)
&=& \sum_{i=1}^n [ D_i \log\{f_1(\bx_i, \by_i, \bvarphi)\}
+(1-D_i) \log\{f_0(\bx_i, \by_i, \bvarphi)\} ] \\
&=& \sum_{i=1}^n [ D_i \log\{ g(\bx_i, \by_i, \bvarphi)\}
+(1-D_i) \log\{1- g(\bx_i, \by_i, \bvarphi)\} ] \\
&&
+ \sum_{i=1}^n \log\{f(\bx_i, \by_i)\}
- n_1\log\left\{ \int g(\bx, \by, \bvarphi)f(\bx, \by)d\bx d\by \right\} \\
&&
- n_0\log\left\{ 1-\int g(\bx, \by, \bvarphi)f(\bx, \by)d\bx d\by \right\}.
\eas
It is necessary to make a direct comparison of the  retrospective likelihood
in \eqref{retro-likelihood}
and  \cite{Lin1997}'s prospective likelihood (defined below),
since the former is   the foundation of this paper
while the latter is the infrastructure in
the variance-component score tests such as SKAT, SKAT-O and MiST.

In the derivation of \eqref{retro-likelihood},  we have assumed that
the random effects $\bv_i$'s for different individuals are independent from each other.
By contrast,  \cite{Lin1997}'s likelihood is derived under the common-random-effect assumption
and regarding  the case-control data  as prospective data.
Under these assumptions, \cite{Lin1997}'s likelihood function is
\ba
\label{lin-likelihood}
L (\bvarphi)
&=&
\int \prod_{i=1}^n    \{\pi_i(\bv, \bvarphi)\}^{D_i}  \{1-\pi_i(\bv, \bvarphi)\}^{1-D_i}   h(\bv) d\bv,
\ea
where $\pi_i(\bv, \bvarphi) = \pi ( \alpha_p +
 \bx_i^\T \bbeta  +  \by_i^\T \bone \gamma   + \by_i^\T  \bv \cdot \sqrt{\theta} )$.
In this model, it seems not possible to find a valid variance estimator
for any point estimators since no replicates from the random effect $\bv$ are available.
In contrast in conventional random effect model, the likelihood is
\[
L_C (\bvarphi)
=
 \prod_{i=1}^n   \int \{\pi_i(\bv, \bvarphi)\}^{D_i}
  \{1-\pi_i(\bv, \bvarphi)\}^{1-D_i}   h(\bv) d\bv.
\]
Unlike $L (\bvarphi)$, here the order of integration and  product is changed.

To test  $H_0: \gamma=0\ \&\  \theta=0$,
it is well accepted that   score tests are preferable to  likelihood ratio tests
 because they avoid using the form of $h(\bv)$.
 Even if $h(\bv)$ is known, estimating  $\theta$ is extremely difficult under alternative hypotheses because
 the integral $\int g(\bx, \by, \bvarphi)f(\bx, \by)d\bx d\by$  involves
a $(d+q)$-dimensional multiple integral,
which is rather complicated particularly  for large  $q$.
Generally the score tests derived from (3) with respect to $\gamma$ and $\theta$
involves many interaction terms between individuals. As a consequence they
are correlated. To simplify their approach,  \cite{Wu2011} removed some terms
in their score equations. It is not clear whether the remaining terms still have a score interpretation
 though the resulting test statistics are valid.
 Moreover, it is not straightforward to combine the two score
 statistics, one for the fixed effect and the other for the random
 effect, to achieve
satisfactory power under different types of alternatives.
To overcome the correlation between the two score statistics,
  \cite{Sun2013} derived a new score statistic with respect to $\theta$
without necessarily restricting  $\gamma=0$.
Their approach may have power loss since their derivations are based on a prospective likelihood
while the available data are retrospectively collected.  In this paper we directly find
the retrospective-likelihood-based score tests
and elucidate their relationship with   \cite{Sun2013}'s score tests.

\subsection{Retrospective score statistic  for $\theta$}
\label{retro-score-theta}

Differentiating the retrospective log-likelihood with respect to $\theta$, we have
the score
\bas
 \left.\frac{\partial \ell_{\retr}}{\partial \theta}\right|_{\theta=0 }
&=&
\left. \sum_{i=1}^n    \frac{D_i-g(\bx_i, \by_i, \bvarphi)}{g(\bx_i, \by_i, \bvarphi) \{1-g(\bx_i, \by_i, \bvarphi)\} }
\frac{\partial  g(\bx_i, \by_i, \bvarphi)}{\partial \theta } \right|_{\theta=0 }\\
&&
+ \left\{  \frac{n_0}{ 1-\int  g(\bx,\by, \bvarphi)   f(\bx,\by)d\bx d\by }
 - \frac{n_1}{ \int g(\bx,\by, \bvarphi) f(\bx,\by)d\bx d\by  }
\right\} \\
&&
\times  \int \left.\frac{\partial
g(\bx, \by, \bvarphi)}{\partial \theta }  f(\bx,\by)d\bx d\by \right|_{\theta=0 }.
\eas
Since   $g(\bx, \by, \bvarphi)$ reduces to
$\pi(\alpha_p+\bbeta^\T \bx +  \by^\T \bone \gamma  )$ when $\theta=0$,
by  L'Hospital's rule, we have
\ba
\left.\frac{\partial  g(\bx, \by, \bvarphi)}{\partial \theta }\right|_{\theta=0}
&=&
\{ 1-2\pi( \alpha_p + \bbeta^\T \bx +  \by^\T \bone \gamma )\}
\pi( \alpha_p + \bbeta^\T \bx +  \by^\T \bone \gamma ) \nonumber \\
&&
\times
\{1-\pi( \alpha_p + \bbeta^\T \bx +  \by^\T \bone \gamma )\}    \by^\T   \by.
\label{partial-g}
\ea
Accordingly we obtain
\bas
&&  \left.\frac{\partial \ell_{\retr}}{\partial \theta}\right|_{\theta=0} \\
&=&
 \sum_{i=1}^n \left\{ D_i - \pi(\alpha_p+  \bbeta^\T \bx_i +  \by_i^\T \bone \gamma  )  \right\}
 \{1-2 \pi(\alpha_p+ \bbeta^\T \bx_i +  \by_i^\T \bone \gamma  )\}     \by_i^\T \by_i  \\
&&
+    n_0  \int  \{1-2 \pi(\alpha_p+  \bbeta^\T \bx +  \by^\T \bone \gamma  )\}
\pi(\alpha_p+ \bbeta^\T \bx +  \by^\T \bone \gamma   )   \by^\T \by   f_0(\bx,\by)d\bx d\by \\
&&
 -  n_1     \int  \{1-2 \pi(\alpha_p+  \bbeta^\T \bx +  \by^\T \bone \gamma )\}
\{  1-\pi(\alpha_p +  \bbeta^\T \bx +  \by^\T \bone \gamma )\} \by^\T \by  f_1 (\bx,\by)d\bx d\by.
\eas

To implement the score test statistic, we need to estimate
the unknown parameters $(\alpha_p, \bbeta, \gamma)$ under $\theta=0$.
Notice that the density functions of the cases and controls
satisfy the density ratio model \citep{Anderson1979, Qin1997}
\bas
f_1(\bx, \by)  = \exp(\alpha_r + \bbeta^\T \bx +  \by^\T \bone \gamma) f_0(\bx, \by),
\eas
where $\alpha_r = \alpha_p +\log\{(1-p)/p\}$  and
 $p =\pr(D=1) = \int f(\bx, \by) g(\bx, \by, \bvarphi)d\bx d\by$   denotes
 the prevalence of the disease of interest.
Throughout this paper, we use $\bbeta$ and $ \alpha$'s, such as $\alpha_r$ and $\alpha_p$,
to denote both argument variables and their true values to save notation;
their meanings are clear from the context.
 Even if $\alpha_r$ is known,  the parameter $\alpha_p = \alpha_r-\log\{ (1-p)/p \}$
is generally unknown because the prevalence $p$ is unknown.
  For the time being, we assume the prevalence $p $ is known, therefore
estimation of $\alpha_p $ is equivalent to the estimation of $\alpha_r$.

 According to   \cite{Qin1997}, we can estimate $( \alpha_r, \bbeta, \gamma)$
 by the maximizer $(\tilde \alpha_r, \tilde \bbeta, \tilde \gamma)$ of
\bas
 \ell_{e,1} (\alpha_r, \bbeta, \gamma)
=
 \sum_{i=1}^n D_i(\alpha_r+\bbeta^\T \bx_i +  \by_i^\T \bone \gamma)
- \sum_{i=1}^n \log\{1+ (n_1/n_0) \exp(\alpha_r +\bbeta^\T \bx_i +  \by_i^\T \bone \gamma)\}.
\eas
Let $F_0(\bx,\by)$ and $F_1(\bx,\by)$ be the true
distribution functions corresponding to  $f_0(\bx,\by, \bvarphi)$ and $f_1(\bx,\by, \bvarphi)$, respectively.
Under  $\theta=0$, the   maximum empirical likelihood estimators of $F_0(\bx,\by)$ and $F_1(\bx,\by)$ are
\bas
\tilde F_0(\bx,\by)
&=& \frac{1}{n_0} \sum_{i=1}^n
\{1-\pi(\tilde \alpha +\tilde \bbeta^{\T} \bx_i
+ \by_i^\T \bone \tilde \gamma )\} I(\bx_i\leq \bx, \by_i\leq \by),  \\
\tilde F_1(\bx,\by)
&=& \frac{1}{n_1} \sum_{i=1}^n  \pi(\tilde \alpha
+\tilde \bbeta^{\T} \bx_i + \by_i^\T \bone \tilde \gamma )  I(\bx_i\leq \bx, \by_i\leq \by),
\eas
where  $\tilde\alpha=\tilde \alpha_r+\log( n_1/n_0)$,
$I(\cdot)$ is the indicator function and the inequality
$\bx_i\leq \bx$ holds elementwise.

Putting these estimators into
${\partial \ell_{\retr}} /\partial \theta |_{\theta=0}$ leads to
\ba
\label{u-alpha}
U_1 (\alpha_p)
&=&
  \sum_{i=1}^n  \{   D_i - \pi(\tilde \alpha +\tilde \bbeta^{\T} \bx_i + \by_i^\T \bone \tilde \gamma ) \}
    \{1-2 \pi(\alpha_p+  \tilde \bbeta^{\T} \bx_i + \by_i^\T \bone \tilde \gamma  )\}  \by_i^\T\by_i,
\ea
which is the retrospective score statistic with respect to $\theta$ if $\alpha_p$ is known.
We shall discuss the case when $\alpha_p$ is unknown later.

\subsection{Retrospective score statistic  for $\gamma$ \label{retros-gamma}}
\label{retro-score-gamma}

Since
\bas
\left.\frac{\partial  g(\bx, \by, \bvarphi)}{\partial \gamma }\right|_{\theta=0, \gamma=0}
&=&
\pi( \alpha_p + \bx^\T \bbeta   )
\{1-\pi( \alpha_p + \bx^\T \bbeta  )\}   \by^\T \bone,
\eas
with similar derivation to the score statistic with respect to $\theta$,
we have  the  retrospective score  with respect to $\gamma$,
\bas
 \left.\frac{\partial \ell_{\retr}}{\partial \gamma}\right|_{\theta=0, \gamma=0}
&=&
 \sum_{i=1}^n   \{D_i-\pi(\alpha_p+  \bx_i^\T \bbeta   ) \} \by_i^\T \bone  \\
 &&
+    n_0    \int \pi( \alpha_p + \bx^\T \bbeta   )   \by^\T \bone  f_1(\bx,\by)d\bx d\by   \\
&&   -  n_1     \int
\{1-\pi( \alpha_p + \bx^\T \bbeta  )\}   \by^\T \bone  f_0(\bx,\by)d\bx d\by.
\eas

To implement the score test statistics,
we need to estimate the unknown parameters $(\alpha_p, \bbeta)$ under $H_0$.
Recall that the density functions of the cases and controls are linked by
\bas
f_1(\bx, \by, \bvarphi)  = \exp(\alpha_r +\bbeta^\T  \bx ) f_0(\bx, \by, \bvarphi).
\eas
According to \cite{Qin1997}, we  estimate $( \alpha_r, \bbeta)$ by the maximizer $(\hat \alpha_r, \hat \bbeta)$ of
\bas
 \ell_1 (\alpha_r, \bbeta)
&=& \sum_{i=1}^n D_i(\alpha_r+\bbeta^{\T} \bx_i) - \sum_{i=1}^n \log\{1+ (n_1/n_0) \exp(\alpha_r +\bbeta^{\T} \bx_i)\}.
\eas
Under  $H_0$, the   maximum empirical likelihood estimators of $F_0(\bx,\by)$ and $F_1(\bx,\by)$ are
\bas
\hat F_0(\bx,\by)
&=& \frac{1}{n_0} \sum_{i=1}^n \{1-\pi(\hat \alpha +\hat \bbeta^{\T} \bx_i )\} I(\bx_i\leq \bx, \by_i\leq \by),  \\
\hat F_1(\bx,\by)
&=& \frac{1}{n_1} \sum_{i=1}^n  \pi(\hat \alpha +\hat \bbeta^{\T} \bx_i )  I(\bx_i\leq \bx, \by_i\leq \by),
\eas
where  $\hat \alpha=\hat \alpha_r+\log( n_1/n_0)$.

Putting $\hat \alpha$, $\hat \bbeta $, $\hat F_0(\bx, \by)$ and $\hat F_1(\bx, \by)$
into $ {\partial \ell_{\retr}}/\partial \gamma |_{\theta=0, \gamma=0}$,
 we have the  retrospective score statistic with respect to $\gamma$,
\ba
\label{u-gamma}
U_2
&=&
  \sum_{i=1}^n  \{   D_i - \pi(\hat \alpha +\hat \bbeta^\T \bx_i) \}     \by_i^\T \bone,
\ea
which   is independent of $\alpha_p$.

\subsection{Prospective score statistic with $\theta$}
\label{prop-score-theta}

If we treat the case-control data as if they were  prospectively collected data,  namely
$\{ (D_i, \bx_i, \by_i): i=1,2, \ldots, n\}$ were
  independent and identically distributed random elements,
the resultant prospective  log-likelihood is
\bas
\ell_{\pros}
&=& \sum_{i=1}^n [ D_i \log\{g( \bx_i, \by_i, \bvarphi)\}
+(1-D_i) \log\{1-g( \bx_i, \by_i, \bvarphi)\} ]
+ \sum_{i=1}^n \log\{f(\bx_i, \by_i)\}.
\eas

The prospective score with respect to $\theta$ without restricting $\gamma=0$ is
\ba
 \left.\frac{\partial \ell_{\pros}}{\partial \theta}\right|_{\theta=0 }
&=&
 \sum_{i=1}^n   \{D_i -\pi(\alpha_p+  \bbeta^\T \bx_{i} + \by_i^\T \bone \gamma)\}
   \{1-2 \pi(\alpha_p+  \bbeta^\T \bx_{i} + \by_i^\T \bone \gamma)\}  \by_i^\T\by_i.
\label{score-pros}
\ea
The  unknown parameters $\alpha_p$, $\bbeta$ and $\gamma$ are
estimated by  the maximum prospective likelihood estimator under $\theta=0$.
The prospective likelihood $\ell_{\pros} $ under $\theta=0$ reduces to
(up to a quantity independent of the parameters)
\ba
  \ell_{e,2} (\alpha_p, \bbeta, \gamma)
&=&  \sum_{i=1}^n D_i(\alpha_p +\bbeta^\T \bx_{i} + \by_i^\T \bone \gamma)
- \sum_{i=1}^n \log\{1+  \exp(\alpha_p  +\bbeta^\T \bx_{i} + \by_i^\T \bone \gamma)\}.
\label{like2}
\ea
Hence the maximum likelihood estimator of $(\alpha_p, \bbeta, \gamma)$ is
\[
(\check{\alpha}_p, \check{\bbeta},  \check{\gamma})
= \arg\max_{\alpha_p, \bbeta, \gamma}   \ell_{e,2} (\alpha_p, \bbeta,\gamma).
\]

If we denote
\(
\alpha=\alpha_r+\log( n_1/n_0) = \alpha_p +\log\{(1-p)/p\}+\log( n_1/n_0)
\),
then
\(
  \ell_{e,1} (\alpha_r, \bbeta,\gamma) =   \ell_{e,2} (\alpha, \bbeta,\gamma) - n_1 \log( n_1/n_0)
\).
Since $\tilde \alpha=\tilde \alpha_r+\log( n_1/n_0)$, it follows that
\bas
(\tilde  \alpha, \tilde  \bbeta, \tilde \gamma)
 =  \arg\max_{\alpha, \bbeta, \gamma}\ell_{e,2} (\alpha, \bbeta,\gamma)
 = \arg\max_{\alpha_p, \bbeta,\gamma}   \ell_{e,2} (\alpha_p, \bbeta,\gamma)
=
(\check{\alpha}_p, \check{\bbeta}, \check{\gamma}).
\eas
When  replacing $(\alpha_p, \bbeta,\gamma)$ by
$(\check{\alpha}_p, \check{\bbeta}, \check{\gamma})
= (\tilde \alpha, \tilde \bbeta, \tilde \gamma )$
in  \eqref{score-pros}, we obtain
 the prospective score statistic with respect  to $\theta$
\ba
&& \sum_{i=1}^n   \{D_i -\pi(\tilde \alpha + \tilde\bbeta^\T \bx_{i} + \by_i^\T \bone \tilde \gamma)\}
  \{1-2 \pi(\tilde \alpha +\tilde\bbeta^\T \bx_{i} + \by_i^\T \bone \tilde \gamma)\}  \by_i^\T\by_i    = U_1(\tilde \alpha),
\label{pros}
\ea
where $ U_1(\cdot)$  is defined in \eqref{u-alpha}.
That is,  the only difference between the prospective score statistic
and the retrospective score statistic with respect  to $\theta$    is   using different $\alpha$ values  in
  $\{1-2 \pi(\alpha +\tilde\bbeta^\T \bx_{i} + \by_i^\T \bone \tilde \gamma) \} \by_i^\T\by_i$.
We shall show that this difference sometimes can lead to severe power loss in hypothesis testing of genetic effect.

\subsection{Prospective score statistic with respect to $\gamma$}
\label{prop-score-gamma}

By direct calculations, we have  the prospective score with respect to $\gamma$
\ba
 \left.\frac{\partial \ell_{\pros}}{\partial \gamma}\right|_{\theta=0, \gamma=0}
&=&
 \sum_{i=1}^n   \{D_i -\pi(\alpha_p+  \bbeta^\T \bx_i)\}    \by_i^\T \bone.
 \label{score-pros-gamma}
\ea
The  unknown parameters $\alpha_p$ and $\bbeta$ are
estimated by  maximizing the prospective likelihood estimator under $H_0$,
which up to a quantity independent of the parameters is
\ba
\label{like2}
  \ell_2 (\alpha_p, \bbeta)
&=&  \sum_{i=1}^n D_i(\alpha_p +\bbeta^{\T} \bx_i) - \sum_{i=1}^n \log\{1+  \exp(\alpha_p  +\bbeta^{\T} \bx_i)\}.
\ea
Hence the maximum likelihood of $(\alpha_p, \bbeta)$ is
given by $(\breve{\alpha}_p, \breve{\bbeta}) = \arg\max_{\alpha_p, \bbeta}   \ell_2 (\alpha_p, \bbeta)$.

Notice that
\(
  \ell_1 (\alpha_r, \bbeta) =   \ell_2 (\alpha, \bbeta) - n_1 \log( n_1/n_0)
\) because
$
\alpha=\alpha_r+\log( n_1/n_0)
$.
Since $\hat \alpha=\hat \alpha_r+\log( n_1/n_0)$, it follows that
\bas
(\hat \alpha, \hat \bbeta) =  \arg\max_{\alpha, \bbeta}\ell_2 (\alpha, \bbeta) = \arg\max_{\alpha_p, \bbeta}   \ell_2 (\alpha_p, \bbeta)
=
(\breve{\alpha}_p, \breve{\bbeta}),
\eas
where $\hat \alpha, \hat \alpha_r$ and $\hat \bbeta$ are defined in Section \ref{retros-gamma}.
When  replacing $(\alpha_p, \bbeta)$ by
$(\breve{\alpha}_p, \breve{\bbeta})= (\hat \alpha, \hat \bbeta )$ in  \eqref{score-pros-gamma},
we can find that the prospective  score statistic with respect to $\gamma$ is exactly equal to
$U_2$, which is the retrospective  score  statistic with respect to $\gamma$.

\section{Asymptotics}
\label{asymptotics}

This section studies the limiting distributions of
the retrospective and prospective score statistics in  \eqref{u-alpha}, \eqref{u-gamma},  and \eqref{pros}
for both retrospectively and prospectively collected data.

\subsection{Asymptotical normality}
\label{asy-normality}

For convenience we assume that  $n_0/n$ is a constant as $n$ goes to infinity, where $n=n_0+n_1$ is the total number
of controls and cases.
Denote   $\bxi_0^\T =(\alpha^\T, \bbeta^\T)$
and  $\bxi_0^\T =(\alpha_p^\T, \bbeta^\T)$, respectively,
as the true parameter values for retrospective and prospective data.
Denote $\bxi_{  *}=(\alpha_*, \bbeta^\T)^\T$,
 $\bz_i = (1, \bx_i^\T)^\T$, $\bz_{e, i}= (1, \bx_i^\T, \by_i^\T \bone)^\T$,
$\bz = (1, \bx^\T)^\T$ and $\bz_e = (1, \bx^\T, \by^\T \bone)^\T$.
Define
\bas
C_2(\bx, \by)
&=&  \by^\T  \bone -  \bS_{xy}^\T \bS_x^{-1} \bz \\
C_1(\bx, \by, \alpha_*)
&=&
\{1-2 \pi(\bxi_*^{\T} \bz_i )\} \by^\T\by -   \bH  (\alpha_*)   \bS_{e,x}^{-1}   \bz_e,
\eas
where
\ba
\label{Sx}
\bS_x  &=&
\frac{1}{n} \e \left[ \sum_{i=1}^n   \pi(\bxi_0^{\T} \bz_i ) \{1- \pi(\bxi_0^{\T} \bz_i ) \}  \bz_i \bz_i^{\T}  \right],
\\
\bS_{e,x}
&=&
\frac{1}{n} \e \left[ \sum_{i=1}^n   \pi(\bxi_{0}^{\T} \bz_{ i} )
\{1- \pi(\bxi_{0}^{\T} \bz_{i} ) \}   \bz_{e,i} \bz_{e,i}^{\T}  \right],
\label{Sx-e}
\\
\label{Sxy}
\bS_{xy}
&=&
\frac{1}{n}\e \left[
\sum_{i=1}^n   \pi( \bxi_0^\T  \bz_i) \{1-\pi(\bxi_0^\T  \bz_i) \}    \by_i^\T \bone   \bz_i
\right],
\\
\label{H}
\bH  (\alpha_*)
&=&
\frac{1}{n} \e
\left[ \sum_{i=1}^n  \{ 1 - \pi(\bxi_{0}^\T \bz_{i}) \}  \pi(\bxi_{0}^\T \bz_{i})
 \{1-2  \pi(\bxi_*^{\T} \bz_i)\}  \by_i^\T\by_i    \bz_{e, i}^\T \right].
\ea
We remark that  all the four quantities are independent of $n$
 and the involved expectation operator $\e$ has different meanings for
 retrospective and prospective data.

If  $(D_i, \bx_i, \by_i)$'s  are case-control or retrospective data,   we define
\bas
\sigma_{11} (\alpha_{1}, \alpha_{2})
&=&
(n_0/n)   \e_0 [  \{\pi(\bxi_0^{\T}   \bz_i ) \}^2   C_1(\bx_i, \by_i, \alpha_{1})  C_1 (\bx_i, \by_i, \alpha_{2}) ]  \\
&&
+
(n_1/n)   \e_1  [ \{1  -   \pi(\bxi_0^{\T}  \bz_i ) \}^2
C_1 (\bx_i, \by_i, \alpha_{1})   C_1 (\bx_i, \by_i, \alpha_{2}) ],
\\
\sigma_{22}
&=&
(n_0/n)   \e_0 [   \{\pi(\bxi_0^{\T}   \bz_i ) \}^2   C_2^2(\bx_i, \by_i )  ]  \\
&&
+
(n_1/n)   \e_1  [ \{1  -   \pi(\bxi_0^{\T}  \bz_i ) \}^2   C_2^2(\bx_i, \by_i )  ].
\eas
where   $\e_0$ ($\e_1$) denotes the expectation operator with respect to  $f_0(\bx, \by, \bvarphi)$ ($f_1(\bx, \by, \bvarphi)$),
where $ \bvarphi$ takes its true value.
If $(D_i, \bx_i, \by_i)$'s  are prospective data,   we define
\bas
\sigma_{11} (\alpha_{1}, \alpha_{2})
&=&
   \e  [  \pi(\bxi_0^{\T}   \bz_i )  \{1  -   \pi(\bxi_0^{\T}  \bz_i ) \}
   C_1(\bx_i, \by_i, \alpha_{1})  C_1 (\bx_i, \by_i, \alpha_{2}) ],
\\
\sigma_{22}
&=&
   \e  [    \pi(\bxi_0^{\T}   \bz_i ) \{1  -   \pi(\bxi_0^{\T}  \bz_i ) \}   C_2^2(\bx_i, \by_i )  ],
\eas
where  $\e $  denotes the expectation operator with respect to  $f (\bx, \by)$.

\begin{theorem}
\label{normality}
Assume $\e (\|\bX\|^2) + \e(\|\bY\|^3) <\infty$ and that $n_1/n $ is a constant as $n$ goes to infinity.
For any  $m$ constants $\alpha_{*1},  \ldots, \alpha_{*m}$,
if $\hat \alpha_{*1},  \ldots, \hat \alpha_{*m}$ are their  consistent estimators,
then  as $n$ goes to infinity,
\[
n^{-1/2}\left( U_{1}(\hat \alpha_{*1}),   \ldots, U_{1}(\hat \alpha_{*m}),  U_{2}\right )^\T
\rightarrow
N\left(0_{(m+1)\times 1},\   \diag( \bSigma(\alpha_{*1}, \ldots, \alpha_{*m}), \sigma_{22} )\right)
\]
in distribution, where
$  \bSigma(\alpha_{*1}, \ldots, \alpha_{*m})
   = \left( \sigma_{11} (\alpha_{*i}, \alpha_{*j}) \right)_{1\leq i, j\leq m}.  $
\end{theorem}

The retrospective and prospective score statistics with respect to $\theta$
are  $U_{1}(\alpha_p)$  and $U_1(\tilde \alpha)$ respectively,
where $\alpha_p$ is assumed to be known and $\tilde \alpha = \alpha+o_p(1)$
with $\alpha = \alpha_p+\log \{ (1-p)n_1/(p n_0)\}$.
Theorem \ref{normality} indicates that  $U_{1}(\hat \alpha_*)$ is asymptotically independent of $U_2$
for any $\alpha_*$ if $\hat \alpha_*=\alpha_*+ o_p(1)$.
It also implies that both  $\sqrt{n}  U_{1}(\alpha_p) $ and $\sqrt{n}  U_{1}(\tilde \alpha ) $
converge  in distribution to  normal distributions with mean zero,
but their asymptotic variances $ \sigma_{11}(\alpha_p, \alpha_p) $ and $ \sigma_{11}(\alpha, \alpha ) $
  are   generally different. If
the proportion $n_1/n$ of the cases in the case-control data is equal
to  the prevalence $p$, then  $\alpha=\alpha_p$ and
the retrospective and prospective score tests and their limiting distributions coincide.

\subsection{Estimation of Variance matrix}
\label{est-var}

Applying the retrospective or prospective tests necessitates consistent estimators
of their corresponding asymptotical variances.
Since $\tilde \bxi$ is  a root-$n$ consistent estimator of $\bxi_0$  whether  the data is retrospective or prospective,
natural root-$n$ consistent estimators of $\bS_x$,  $\bS_{e,x}$,   $\bS_{xy}$ and $\bH(\alpha_*) $ are
\ba
\label{hatsx}
\tilde \bS_x
&=&
\frac{1}{n} \sum_{i=1}^n      \pi(\tilde \bxi^{\T} \bz_{i}  )
\{1- \pi(\tilde \bxi^{\T} \bz_{i}  ) \}     \bz_i \bz_i^{\T},
\\
\label{hatsx-e}
\tilde \bS_{e,x}
&=&
\frac{1}{n} \sum_{i=1}^n      \pi(\tilde \bxi^{\T} \bz_{i}  )
\{1- \pi(\tilde \bxi^{\T} \bz_{i}  ) \}     \bz_{e,i} \bz_{e,i}^{\T},
\\
\label{hatsxy}
\tilde \bS_{xy}
&=&
\frac{1}{n}
\sum_{i=1}^n   \pi( \tilde \bxi^{\T} \bz_{i}  ) \{1-\pi(  \tilde \bxi^{\T} \bz_{i}  ) \}    \by_i^\T \bone   \bz_i,    \\
\label{hatH}
\tilde \bH  ( \alpha_*)
&=&  \frac{1}{n} \sum_{i=1}^n  \{ 1 - \pi( \tilde \bxi^{\T} \bz_{i} ) \}  \pi(\tilde \bxi^{\T} \bz_{i} )
 \{1-2  \pi(  \tilde \bxi^\T_*   \bz_{i})\} \by_i^\T\by_i   \bz_{e,i}^\T,
\ea
where $\tilde \bxi_* = (\alpha_*, \tilde \bbeta)$.
Further, define
\ba
\hat \sigma_{11} (  \alpha_{*1},   \alpha_{*2})
&=&
\frac{1}{n}  \sum_{i=1}^n  \pi(\tilde \bxi^\T   \bz_{i} ) \{ 1- \pi(\tilde \bxi^\T   \bz_{i}) \}
 \tilde C_1 (\bx_i, \by_i, \alpha_{*1}) \tilde C_1 (\bx_i, \by_i, \alpha_{*2}),
\label{sigma11hat}
\\
\hat \sigma_{22}
&=&
\frac{1}{n}  \sum_{i=1}^n  \pi(\tilde \bxi^\T   \bz_{i} ) \{ 1- \pi(\tilde \bxi^\T   \bz_{i}) \}
\{ \tilde C_2^\T(\bx_i, \by_i) \}^2,
\label{sigma22hat}
\ea
where
$
\tilde C_1(\bx, \by, \alpha_*)
=
\{1-2 \pi(\tilde \bxi_{*}^\T   \bz )\} \by^\T\by -   \tilde \bH  (  \alpha_*)  \tilde S_{e,x}^{-1} \bz_e
$
and
$
\tilde C_2(\bx, \by)
=
 \by^\T  \bone -  \tilde \bS_{xy}^\T \tilde \bS_x^{-1}   \bz.
$
One can straightforwardly verify  that for any two constants $\alpha_{*1}$ and $\alpha_{*2}$,
 if $\hat \alpha_{*1}$ and $\hat \alpha_{*2}$ are their consistent estimator, then
$\hat \sigma_{11} (\hat \alpha_{*1},  \hat \alpha_{*2})$
and
$\hat \sigma_{22} $
are  consistent estimators of
$ \sigma_{11} ( \alpha_{*1},   \alpha_{*2})$
and
$ \sigma_{22} $, respectively,
whether the data is retrospective or prospective.

\section{Proposed score tests}
\label{proposal}

Denote the standardized score tests for random and fixed effects by
\bas
U_{1s}(\alpha_*) =  \frac{ U_1(\alpha_*) /\sqrt{n} }{\sqrt{  \hat \sigma_{11 }(  \alpha_{*},  \alpha_{*}) }}  \quad
\mbox{and} \quad
U_{2s} = \frac{U_2/\sqrt{n}}{\sqrt{ \hat \sigma_{22}}}.
\eas
Theorem \ref{normality} and the consistency of the variance estimators imply that
both $U_{1s}(\alpha_*)$ (for  fixed $\alpha_*$) and $U_{2s}$
are  asymptotically independent and both
have an asymptotically standard normal distribution.
Throughout this section,  all limits are taken under  $H_0: \theta=0 \  \&\  \gamma=0$.
Since  the hypothesis with respect to $\gamma$ is two-sided,
we shall reject $\gamma=0$ if
\[
\fixedsc =   U_{2s}^2
\]  is large enough.
While the hypothesis for $\theta$ is one-sided ($\theta\geq0$) and
a larger  $U_{1s}(\alpha_*)$  supports   $\theta>0$,
we   should  reject $\theta=0$ for large values of
\[
\randomsc(\alpha_*)= \{ U_{1s}^+(\alpha_*) \}^2,
\]   where
$ U_{1s}^+(\alpha_*) = \max\{U_{1s}(\alpha_*), 0\}$.
To capture  non-null hypothesis in both fixed and random effects,
we take both  scores  into account and define
\bas
\sumsc(\alpha_*)
= \{ U_{1s}^+(\alpha_*) \}^2 + U_{2s}^2.
\eas
It is worth pointing out that
the  tests  $\sumsc(\hat \alpha)$ and $\randomsc(\hat \alpha)$ correspond to the prospective score tests,
and
$\sumsc( \alpha_p)$ and $\randomsc( \alpha_p)$     correspond to the retrospective   score tests.
As $n$ goes to infinity,  the limiting distributions of
$\fixedsc, \randomsc(\alpha_*)$ and $\sumsc(\alpha_*)$
are $ \chi_{1}^2$, $0.5\chi_0^2 + 0.5\chi_1^2$ and $0.5\chi_1^2 + 0.5\chi_2^2$, respectively.
Moreover, our combination of score statistics differs from the conventional linear combination of score tests with normally
distributed limiting distributions.

The parameter $\alpha_p$ is generally unknown in practice and the case-control data contains no information about $p$
or $\alpha_p$.
If we have some guesses for $\alpha_p$ such as  $ \alpha_{*i} $ ($i=1,\ldots, m$),
we may define another two tests
\bas
\sumsc(\alpha_{*1}, \ldots, \alpha_{*m}) =
  \max_{1\leq i\leq m} \sumsc(\alpha_{*i}),   \quad
\randomsc(\alpha_{*1}, \ldots, \alpha_{*m})
=  \max_{1\leq i\leq m} \randomsc(\alpha_{*i}).
\eas
By Theorem 1,  $\randomsc(\alpha_{*1}, \ldots, \alpha_{*m})$
converges in distribution to  $\max_{1\leq i\leq m} (Z_i^+)^2$,
where $(Z_1, Z_2, \ldots, Z_m)$
follows a $m$-variate  normal distribution with mean zero and variance
\[
\bSigma_s(\alpha_{*1},  \ldots, \alpha_{*m}) = \left(
  \frac{\sigma_{11 }(  \alpha_{*i},  \alpha_{*j})}{\sqrt{\sigma_{11 }(  \alpha_{*i},  \alpha_{*i})  \sigma_{11 }(  \alpha_{*j},  \alpha_{*j})}}
 \right)_{1\leq i, j\leq m}.
\]
If we denote   $F((t_1,\ldots, t_m), \bSigma)$  the distribution function
of the $m$-dimensional normal distribution with mean zero and variance $\bSigma$,
it follows that    as $n\rightarrow \infty$,
\[
\lim_{n\rightarrow \infty}P( \randomsc(\alpha_{*1}, \ldots, \alpha_{*m}) \leq t ) =
F\left(\sqrt{t} \bone, \; \bSigma_s(\alpha_{*1},  \ldots, \alpha_{*m}) \right).
\]
The distribution   $F((t_1,\ldots, t_m), \bSigma)$ can be calculated by  the  {\tt pmvnorm} function
of the {\tt R} package {\tt mvtnorm}.
Similarly,  $\sumsc(\alpha_{*1}, \ldots, \alpha_{*m})$ converges in distribution to
$Z_{0}^2+ \max_{1\leq i\leq m} (Z_i^+)^2$,  where
 $Z_{0}$ denotes a random variable following the standard normal distribution  and  independent of  $Z_i$ ($1\leq i\leq k$).
Straightforward calculus gives
\[
\lim_{n\rightarrow \infty}P( \sumsc(\alpha_{*1}, \ldots, \alpha_{*m}) \leq t ) =
2\int_0^{\sqrt{t}}
F\left(\sqrt{t-v^2} \bone, \; \bSigma_s(\alpha_{*1},  \ldots, \alpha_{*m}) \right) \phi(v) dv.
\]

In practice, we may make an interval guess about the prevalence $p$ by experience or prior information.
In cancer studies, this information can be easily retrieved through\\ 
 \centerline{ \url{https://seer.cancer.gov/data/},}\\
the Surveillance, Epidemiology, and End Results Program (SEER), an authoritative source for cancer statistics in the US.
Let $[b_1, b_2]$ be  an interval guess for   $\log\{p/(1-p)\}$  and let $\hat \alpha$  be
the maximum empirical likelihood estimator of $\alpha$, which together with $\hat \bbeta$ maximizes \eqref{like2}.
Given  $m>1$, we propose to choose $\alpha_{*i}=\hat \alpha - \log(n_1/n_0) + (i-1)\times (b_2-b_1)/(m-1)$
and define
\bas
\randomsc( [b_1, b_2], m)
= \randomsc(\alpha_{*1}, \ldots, \alpha_{*m}), \quad
\sumsc( [b_1, b_2], m)
= \sumsc(\alpha_{*1}, \ldots, \alpha_{*m}) .
\eas
In our simulation study, we have chosen $m=4$ and $[b_1, b_2]=[-10, -0.5]$,  which
corresponds to the case that the disease prevalence $p$ falls in the interval $[4.54\times 10^{-5},\;  0.38]$.

\section{Simulation}
\label{simulation}

We conduct simulations to  investigate  the finite-sample performance
(including type I error and power) of the proposed tests.
We simulated case-control data with an equal number $n_0=n_1=2000$ of
cases and controls from
\bas
 \pr(D=1|\bx, \by, \bv)
 = \pi\left(\alpha_p + \bx^\T \bbeta
  +  \by^\T \bgamma +  \by^\T  \bv \cdot \sqrt{\theta}  \right).
\eas
The covariate is $\bx=(x_1, x_2)^\T$, where  $x_1$ and $x_2$  are
independently generated from a Bernoulli distribution with success
probability 0.5 and   $N(1, 1)$, respectively.
The components of the random effect $\bv$ are independent and identically distributed as N(0,  1).
The genotype values were simulated under Hardy-Weinberg equilibrium and linkage equilibrium.
In the simulation studies below, the minor allele frequency (MAF) is considered in all possible ranges, from common to rare.
The MAF refers to the frequency at which the second most common allele occurs in a given population.
More details can be found through\\
\centerline{\url{https://en.wikipedia.org/wiki/Minor_-allele_-frequency}.}\\
All the reported numbers in this section were calculated based on  2000 simulation repetitions.

Our simulation settings mimic those of  \cite{Sun2013}  but  with  two differences.
One is that \cite{Sun2013}  considered continuous covariate case only,
but our data generating model involves both  binary and  continuous covariates.
 This difference is not essential.  The other is that
 all subjects in a simulated study share the same random effect
 in \cite{Sun2013}  if it does exist,
but in our settings the random effects are independent and identically
distributed across all individuals.

\subsection{Test for random effect}
We first consider the problem of testing the existence of random effect, namely $H_0: \theta = 0$.
The proposed random effect score test  $\randomsc (\alpha_*)$
is designed for this purpose.
We investigate the type I error and power of $\randomsc (\alpha_*)$
with \( \alpha_* = \alpha_p  \)  or $ \hat \alpha $,  and $\randomsc$--MAX
by comparing them with \cite{Sun2013}'s random effect score test (MiST-Random for short),
which is available from the R package {\tt MiST}.
Among the existing tests for genetic effect,
\cite{Sun2013}'s MiST-Random test is the only  one  designed for testing  the existence of random effect.

\begin{example}
\label{ex-a}
When generating the genotype  vector  $\by $,
the MAFs of the genotypes are set to  ${\rm MAF}_j =  j/(3q+1)$ ($j=1,2,\ldots, q$).
We set the dimension of $\bv$ and $\by$ to   $q=10$,
and set  $\alpha_p=-1$ and $\bbeta=(0.5,-1)^\T$.
We consider  four scenarios:
(C1)  $\sqrt{\theta}= 0 $, $\bgamma=  (-0.02k)\times \bone_q $;
(C2)  $\sqrt{\theta}= 0.5 $, $\bgamma=   (-0.02k)\times \bone_q $;
(C3) $\sqrt{\theta} = 0.15k$, $\bgamma=  0\times {\bf 1}$;
(C4) $\sqrt{\theta}= 0 $, $\bgamma= (0.05k)\times \bgamma_{0} $,
where
$\bgamma_{0} =(\bone_{q/2}^\T,  \;  -\bone_{q/2}^\T)^\T$ and $k=0, 1, \ldots, 5$.
The disease prevalence is between 0.15 and 0.25.
\end{example}

Table  \ref{case-ex-a1} tabulates the  rejection rates of
$\randomsc (\alpha_p)$, $\randomsc (\hat \alpha)$, RS-MAX and MiST-Random.
Model \eqref{random-effect-model} is correct in the first three scenarios.
In scenario (C1),   the null hypothesis  $H_0: \theta = 0$ is  true and
the reported numbers for this scenario  are type I errors.
We see that all  the four tests under consideration have desirable control on their type I errors.
In scenario (C2),   the null hypothesis  $H_0: \theta = 0$ is violated, hence
the corresponding  numbers for this scenario  are powers.
It seems that the MiST--Random test has no power at all
while the proposed $\randomsc$ tests have desirable power.
In scenario (C3),  the true model is getting farther and farther away from the null as $k$
increases from 0 to 5.   The power of the proposed $\randomsc$ tests are still
desirable  but   the MiST--Random test  has no power again.
These indicates that when  model \eqref{random-effect-model} is correct,
although the MiST--Random test can control its type I error,  it may lose power remarkably.
By contrast, the proposed $\randomsc$ tests have much better power when the null is violated
and much better  type I error when the null is true.
Scenario (C4) represents the case where the null hypothesis $H_0: \theta=0$ is true but
model \eqref{random-effect-model} is not rigorously correct.
In this scenario, the proposed $\randomsc$ tests still can control their type I errors very well,
however  the MiST--Random test has severely inflated type I errors, which can be   as large as 100\%.

\begin{table}
\centering
\caption{ \large
Rejection probabilities (\%) of the four tests for random effect
in Scenarios (C1)-(C4) of Example \ref{ex-a} ($\bgamma_{0} =(\bone_{q/2}^\T,  \;  -\bone_{q/2}^\T)^\T$)
\label{case-ex-a1}
}
\begin{tabular}{ l  rrrrrr  }
\hline
  &  $k=0$&$k=1$ &$k=2$ &$k=3$ &   $k=4$&$k=5$   \\\hline
                         &     \multicolumn{6}{c }{ (C1)  $\sqrt{\theta}= 0 $, $\bgamma=  (-0.02k)\times \bone_q $ }   \\

$\randomsc$--MAX           &  3.85&  4.75&  4.60&  4.95&  7.30&  5.55     \\
$\randomsc(\alpha_p)$      &  2.80&  4.05&  4.90&  4.40&  7.40&  5.75     \\
$\randomsc(\hat \alpha )$  &  4.60&  6.15&  4.40&  4.85&  4.85&  5.35     \\
MiST--Random               &  4.55&  5.20&  4.90&  4.95&  5.10&  4.85     \\\hline

                         &     \multicolumn{6}{c }{ (C2)  $\sqrt{\theta}= 0.5 $, $\bgamma=  (-0.02k)\times \bone_q $ }    \\

$\randomsc$--MAX           &  60.40& 55.85& 59.15& 59.80& 65.00& 68.55    \\
$\randomsc(\alpha_p)$      &  66.10& 64.50& 68.50& 66.25& 72.75& 73.60    \\
$\randomsc(\hat \alpha )$  &  54.35& 49.00& 45.70& 52.60& 46.95& 55.50    \\
MiST--Random               &   5.35&  5.55&  4.85&  4.85&  4.65&  5.60    \\\hline

                         &     \multicolumn{6}{c }{(C3)  $\sqrt{\theta} =  0.15 k$, $\bgamma=  0\times \bone_q$}  \\

$\randomsc$--MAX           & 3.70&  8.50& 20.05& 53.85& 75.85& 87.70   \\
$\randomsc(\alpha_p)$      & 3.60&  9.55& 22.35& 60.90& 83.15& 91.45   \\
$\randomsc(\hat \alpha )$  & 5.30&  9.25& 16.35& 44.70& 67.80& 82.25   \\
MiST--Random               & 4.80&  5.00&  4.65&  5.05&  4.95&  5.90   \\\hline

                         &      \multicolumn{6}{c}{(C4)  $\sqrt{\theta}= 0 $, $\bgamma= (0.05k)\times \bgamma_{0} $}  \\

$\randomsc$--MAX           & 4.80&  2.65&  1.70&  1.00&   1.85&   0.50   \\
$\randomsc(\alpha_p)$      & 4.10&  2.15&  0.60&  0.60&   0.30&   0.00   \\
$\randomsc(\hat \alpha )$  & 4.85&  6.50&  4.65&  3.20&   6.85&   5.20   \\
MiST--Random               & 4.50& 23.45& 82.55& 99.90& 100.00& 100.00   \\\hline
\end{tabular}
\end{table}

\subsection{Test for overall genetic effect}

Next, we consider testing the overall genetic effect, for which  the SS test is designed.
We study the finite-sample performance of the proposed SS test by comparing the following tests:
(1)  the  burden test (Burden for short) calculated by the {\tt R} package {\tt SKAT},
(2) the Sequence Kernel Association Test \cite[SKAT]{Wu2011},
(3)  the optimal test in an extended family of SKAT tests  \cite[SKAT-O]{Lee2012b},
(4)  \cite{Sun2013}'s MiST test,
(5)  $\sumsc (\alpha_p)$,
(6)  $\sumsc (\hat \alpha)$,
and
(7)  $\sumsc$--MAX: \; $\sumsc([-10, -0.5], 4)$.
The simulated rejection rates of these seven tests are reported in Table  \ref{case-ex-a2}.

\begin{table}[!ht]
\def~{\hphantom{0}}
\caption{
Rejection probabilities (\%) of the seven tests for overall genetic effect
and the FS score test in Scenarios (C1)-(C4) of Example \ref{ex-a}
 ($\bgamma_{0} =(\bone_{q/2}^\T,  \;  -\bone_{q/2}^\T)^\T$)
\label{case-ex-a2}}
\centering
\begin{tabular}{ l rrrrrr  }\hline
      &   $k=0$&$k=1$ &$k=2$ &$k=3$ &   $k=4$&$k=5$  \\\hline
                         &     \multicolumn{6}{c }{(C1) $\sqrt{\theta}= 0 $, $\bgamma=  (-0.02k)\times \bone_q $ }     \\

$\sumsc$--MAX              & 91.85& 78.45& 65.85& 50.70& 53.50& 60.65    \\
$\sumsc(\alpha_p)$         & 91.95& 80.85& 70.80& 53.40& 59.05& 63.35    \\
$\sumsc(\hat \alpha )$     & 90.05& 73.35& 55.25& 40.80& 33.00& 47.05     \\
Burden                     & 12.30&  8.05&  5.35&  4.40&  5.40&  8.35    \\
SKAT                       &  7.30&  5.45&  4.55&  5.00&  5.55&  6.10    \\
SKAT-O                     & 11.20&  7.75&  5.35&  4.90&  5.65&  7.85    \\
MiST                       & 70.65& 45.85& 24.55&  6.10&  5.20& 10.95    \\\hline

                         &     \multicolumn{6}{c }{(C2)  $\sqrt{\theta}= 0.5 $, $\bgamma=   (-0.02k)\times \bone_q $ }     \\
$\sumsc$--MAX              &  4.30& 13.30& 38.85& 71.05& 92.80& 99.40    \\
$\sumsc(\alpha_p)$         &  3.95& 14.85& 42.15& 74.40& 94.25& 99.55    \\
$\sumsc(\hat \alpha )$     &  4.45& 17.55& 42.00& 74.20& 93.75& 99.60     \\
Burden                     &  5.85&  5.00& 11.40& 17.20& 28.50& 37.65    \\
SKAT                       &  5.05&  4.40&  6.75&  7.90& 11.80& 14.70    \\
SKAT-O                     &  5.70&  5.15&  9.75& 13.85& 25.15& 32.00    \\
MiST                       &  4.50& 14.90& 38.70& 68.85& 91.85& 99.20    \\\hline

                          &     \multicolumn{6}{c }{(C3)  $\sqrt{\theta} =  0.15k$, $\bgamma=  0\times\bone_q$}  \\

$\sumsc$--MAX             &  5.95&  7.40& 32.40& 62.95&  78.65&  94.00     \\
$\sumsc(\alpha_p)$        &  6.30&  8.55& 36.35& 68.30&  82.65&  94.95     \\
$\sumsc(\hat \alpha )$    &  5.05& 11.85& 39.00& 70.10&  85.20&  95.90     \\
Burden                    &  5.20& 13.30& 34.00& 61.75&  85.00&  95.50     \\
SKAT                      &  5.85&  8.55& 15.45& 31.70&  51.15&  75.05     \\
SKAT-O                    &  6.10& 11.30& 29.50& 57.00&  81.40&  94.10     \\
MiST                      &  5.10& 23.30& 89.00& 99.95& 100.00& 100.00     \\\hline

                          &     \multicolumn{6}{c }{(C4)  $\sqrt{\theta}= 0 $, $\bgamma= (0.05k)\times \bgamma_{0} $} \\

$\sumsc$--MAX             & 4.35&  8.65& 37.45& 89.95& 98.60& 99.80    \\
$\sumsc(\alpha_p)$        & 4.30&  9.65& 40.00& 91.05& 99.00& 99.85    \\
$\sumsc(\hat \alpha )$    & 5.05&  9.85& 35.95& 88.60& 98.05& 99.70    \\
Burden                    & 3.90&  5.80&  7.40& 14.45& 20.80& 23.10    \\
SKAT                      & 4.95&  4.00&  5.45&  6.45&  9.90& 11.45    \\
SKAT-O                    & 4.55&  5.35&  7.40& 12.30& 18.20& 20.55    \\
MiST                      & 5.15&  6.80& 24.95& 73.50& 89.55& 97.45    \\\hline

\end{tabular}
\end{table}

As mentioned previously,  model \eqref{random-effect-model} is correct in the first three scenarios.
In scenario (C1), when $k=0$, the null hypothesis is true  and all the seven tests
have good control on their type I errors.  As $k$ increases from 1 to 5,
the proposed \sumsc tests including  $\sumsc (\alpha_p)$, $\sumsc (\hat \alpha)$, and $\sumsc$--MAX
and the MiST test are almost the same powerful, which
are clearly much more powerful than those of the rest three tests.
In scenario (C2), since $\theta\neq 0$,  the null hypothesis is always violated
regardless of $k$
and the reported rejection probabilities are all  powers.
We can see clearly that  the proposed three \sumsc  tests
outperform the rest four tests by a large margin in terms of power.
In scenario (C3), again the proposed three  \sumsc  tests have very large power gains against
the rest four tests.
In scenario (C4),  Model \eqref{random-effect-model} is violated
and there is no random effect but fixed effect.
As demonstrated in Table  \ref{case-ex-a1},  The MiST-Random test has uncontrolable type I errors.
which means that the powers of MiST are unreliable.
Under this setting, the proposed three \sumsc tests are almost the same powerful
as burden and SKAT-O, and  are more powerful  than SKAT.

It is worth mentioning that in scenario (C2), $\sumsc (\alpha_p)$,
the \sumsc test with the true prevalence,  is much more powerful than
$\sumsc (\hat \alpha)$, which is based on the score tests from the prospective likelihood.
The only difference between the two tests is that
they are using different values of $\alpha_*$ in the proposed SS test $\sumsc (\alpha_*)$.
Note that
\[
\hat \alpha = \alpha_p + \log\{(1-p)/p\} - \log\{ (1-n_1/n)/(n_1/n) \} + o_p(1)
\]
is non-consistent for and  different from $\alpha_p$  unless $n_1/n = p + o_p(1)$.
The substantially different performance of $\sumsc (\alpha_p)$ and $\sumsc (\hat \alpha)$
is therefore because
 the disease prevalence $p$ (between 0.15 and 0.25)  is much different from $n_1/n = 0.5$.
Since $\alpha_p$ is  estimable from  case-control data if the prevalence is known,
the larger powers of $\sumsc (\alpha_p)$  indicate that knowledge of disease prevalence can indeed
be used to enhance overall statistic test power.

\subsection{Low-prevalence scenarios }

The disease prevalence in the scenarios of Example \ref{ex-a}
ranges between 15\% and 25\%, it reflects the common disease case.
Since the case-control design is especially convenient and cost-effective
for a rare disease,  it is natural to see how the proposed tests perform
when the disease prevalence is low, such as around 5\%.

\begin{example}
\label{ex-b}
When generating the genotype  vector  $\by $,
the MAFs of the genotypes are set to  ${\rm MAF}_j =  j/(3q+1)$ ($j=1,2,\ldots, q$).
We set the dimension of $\bv$ and $\by$ to   $q=10$,
and set  $\alpha_p=-2$, $\bbeta=(-1,-1)^\T$.
We consider the following four scenarios:
(D1)  $\sqrt{\theta}= 0 $, $\bgamma=  (-0.015k)\times \bone_q $;
(D2) $\sqrt{\theta}= 0.36 $, $\bgamma=  (-0.015k)\times \bone_q $;
(D3) $\sqrt{\theta} =  0.08k$, $\bgamma=  0\times {\bf 1}$;
(D4) $\sqrt{\theta}= 0 $, $\bgamma=  (0.05k)\times \bgamma_{0} $.
The disease prevalence  is  around 5\% in all scenarios.
\end{example}

The simulated rejection rates are reported in Tables  \ref{case-ex-b1} and  \ref{case-ex-b2}.
The four scenarios in Example \ref{ex-b} are very similar to those in Example \ref{ex-a},
except that the prevalence is  relatively low.
Our observations from Tables  \ref{case-ex-b1} and  \ref{case-ex-b2}
are similar to the previous one in which the proposed tests have desirable controls on type I error in all scenarios
no matter the problem of interest is to test the existence of random effect or to test the overall genetic effect.
When model \eqref{random-effect-model}  is correct and there exists no random effect,
such as in scenario (D1),  the proposed three \sumsc tests clearly have very close performance as MiST,
and all of them outperform Burden, SKAT and SKAT-O by a large margin.
When model \eqref{random-effect-model}  is correct and  random effect does exist,
such as in Scenario (D2),
the proposed  \sumsc test with the true prevalence is the most powerful one
and the data-adaptive \sumsc test, SS-MAX, has very similar power.
These two tests   outperform the rest five tests for overall genetic effect
by a large margin.  This also indicates that the prevalence information
can lead to much power gain.
This observation is still true in scenario (D3) where
model \eqref{random-effect-model}  is correct and there is no fixed effect,
although the advantage of the proposed  \sumsc test with the true prevalence
and the  SS-MAX diminishes.
When model \eqref{random-effect-model}  is violated such as in Scenario (D4)
the proposed tests have less power than MiST. In this situation, however, this is not a fair comparison
because MiST-random loses control for its type I error for testing the existence of random effect.

\begin{table}[!ht]
\caption{
Rejection probabilities (\%) of the  tests under consideration
and the FS score test
for Scenarios (D1)-(D2) in Example \ref{ex-b}
($\bgamma_{0} =(\bone_{q/2}^\T,  \;  -\bone_{q/2}^\T)^\T$)
\label{case-ex-b1}}

\centering
\begin{tabular}{ l  rrrrrr  }\hline
  &  $k=0$&$k=1$ &$k=2$ &$k=3$ &   $k=4$&$k=5$    \\\hline
                           &  \multicolumn{6}{c }{(D1) $\sqrt{\theta}= 0 $, $\bgamma=  (-0.015k)\times \bone_q $ }     \\

$\randomsc$--MAX           & 3.35 & 3.40&  3.15&  3.75&  4.50&  4.55   \\
$\randomsc(\alpha_p)$      & 3.05 & 3.90&  3.55&  4.35&  5.20&  4.70   \\
$\randomsc(\hat \alpha )$  & 3.15 & 2.75&  2.50&  2.60&  2.35&  3.70   \\
MiST--Random               & 4.40 & 4.30&  3.95&  4.60&  3.90&  4.55   \\[1ex]
$\sumsc$--MAX              & 3.50 & 7.35& 19.05& 42.65& 70.25& 86.05   \\
$\sumsc(\alpha_p)$         & 4.10 & 8.50& 22.15& 46.75& 74.40& 89.00   \\
$\sumsc(\hat \alpha )$     & 4.05 & 8.15& 21.75& 45.50& 73.75& 88.00   \\

Burden                     & 5.90 & 4.45&  5.50&  6.95&  9.95& 15.55   \\
SKAT                       & 5.35 & 3.60&  5.25&  5.35&  5.35&  6.50   \\
SKAT-O                     & 5.80 & 4.15&  5.70&  7.15&  8.85& 13.10   \\
MiST                       & 4.55 & 8.55& 19.10& 42.35& 69.85& 85.85   \\ \hline

                        &      \multicolumn{6}{c }{(D2) $\sqrt{\theta}= 0.36 $, $\bgamma= (-0.015k)\times \bone_q $  }     \\

$\randomsc$--MAX           & 32.50& 32.50& 35.45& 33.85& 36.15& 33.15    \\
$\randomsc(\alpha_p)$      & 40.20& 39.70& 40.60& 41.15& 43.65& 40.10    \\
$\randomsc(\hat \alpha )$  & 11.20&  9.60& 10.75&  9.15& 10.25&  9.70    \\
MiST--Random               &  5.35&  4.10&  4.20&  5.20&  4.55&  4.70    \\[1ex]
$\sumsc$--MAX              & 87.75& 77.35& 61.05& 45.70& 31.05& 25.40    \\
$\sumsc(\alpha_p)$         & 89.90& 80.15& 64.60& 48.85& 35.70& 27.65    \\
$\sumsc(\hat \alpha )$     & 83.95& 67.60& 47.70& 26.90& 10.95&  7.15    \\

Burden                     & 23.50& 17.65& 12.00& 10.15&  5.35&  6.00    \\
SKAT                       & 11.90&  9.75&  8.25&  7.85&  6.00&  4.85    \\
SKAT-O                     & 20.20& 15.15& 10.80&  9.70&  5.65&  4.90    \\
MiST                       & 78.40& 60.50& 38.85& 21.15&  8.25&  4.40    \\\hline
\end{tabular}

\end{table}

\begin{table}[!ht]
\caption{
Rejection probabilities (\%) of the  tests under consideration
and the FS score test
for Scenarios (D3)-(D4) in Example \ref{ex-b}
($\bgamma_{0} =(\bone_{q/2}^\T,  \;  -\bone_{q/2}^\T)^\T$)
\label{case-ex-b2}}

\centering
\begin{tabular}{ l  rrrrrr  }\hline
  &  $k=0$&$k=1$ &$k=2$ &$k=3$ &   $k=4$&$k=5$    \\\hline
                         &     \multicolumn{6}{c }{(D3) $\sqrt{\theta} =  0.08k $, $\bgamma=  0\times \bone_q$}   \\

$\randomsc$--MAX           & 3.35& 3.80& 5.60& 11.25& 21.80& 43.60      \\
$\randomsc(\alpha_p)$      & 3.05& 3.50& 6.50& 14.50& 27.80& 50.30      \\
$\randomsc(\hat \alpha )$  & 3.15& 3.10& 3.60&  3.50&  7.80& 14.95      \\
MiST--Random               & 4.40& 5.00& 3.60&  4.45&  4.50&  4.10       \\[1ex]

$\sumsc$--MAX              & 3.50& 3.85& 7.60& 30.90& 73.85& 97.30      \\
$\sumsc(\alpha_p)$         & 4.10& 4.05& 7.95& 34.10& 75.95& 97.90      \\
$\sumsc(\hat \alpha )$     & 4.05& 3.90& 6.55& 25.05& 68.35& 95.95      \\
Burden                     & 5.90& 6.35& 7.25& 12.10& 19.15& 26.15      \\
SKAT                       & 5.35& 6.25& 5.70&  7.60& 10.25& 11.00      \\
SKAT-O                     & 5.80& 6.60& 6.80& 11.20& 16.60& 22.65      \\
MiST                       & 4.55& 4.80& 6.35& 24.85& 62.85& 93.10      \\ \hline

                         &       \multicolumn{6}{c }{(D4) $\sqrt{\theta}= 0 $, $\bgamma= (0.05k)\times \bgamma_{0} $} \\

$\randomsc$--MAX           & 3.35&  0.95&  1.05&  0.95&   1.05&   0.90    \\
$\randomsc(\alpha_p)$      & 3.05&  1.25&  0.45&  0.15&   0.15&   0.00    \\
$\randomsc(\hat \alpha )$  & 3.15&  1.65&  3.05&  2.60&   3.45&   4.40    \\
MiST--Random               & 4.40& 24.50& 84.25& 99.75& 100.00& 100.00     \\[1ex]

$\sumsc$--MAX              & 3.50&  7.95& 24.75& 50.85&  74.35&  91.25    \\
$\sumsc(\alpha_p)$         & 4.10& 10.40& 29.10& 56.65&  78.60&  93.65    \\
$\sumsc(\hat \alpha )$     & 4.05& 10.65& 31.45& 59.75&  80.60&  94.15    \\
Burden                     & 5.90& 14.90& 33.55& 64.50&  89.45&  96.80    \\
SKAT                       & 5.35&  7.65& 16.10& 32.80&  61.70&  82.15    \\
SKAT-O                     & 5.80& 13.30& 29.30& 60.20&  86.20&  96.10    \\
MiST                       & 4.55& 26.85& 88.50& 99.95& 100.00& 100.00    \\\hline
\end{tabular}

\end{table}

\subsection{Rare variants \& low prevalence }

In the previous simulation settings,   the MAFs of the genotypes vary from
3\% to  33\%. Since rare variant generally refers to a variant with MAF no larger than 1\%,
we  consider some  simulation settings in which the MAFs are no larger than 1\%.

\begin{example}
\label{ex-c}
When generating the genotype  vector  $\by $,
the MAFs of the genotypes are set to  ${\rm MAF}_j =  0.005 + 0.005*j/q$ ($j=1,2,\ldots, q$).
We set the dimension of $\bv$ and $\by$ to   $q=10$,
and set  $\alpha_p=-2$, $\bbeta=(-1,-1)^\T$.
We consider the following  scenarios:
(E1) $\sqrt{\theta}= 0 $, $\bgamma=   (-0.08k)\times \bone_q $;
(E2) $\sqrt{\theta}= 0.8+  0.3k$, $\bgamma=  (-0.25k)\times \bone_q $;
(E3) $\sqrt{\theta} = 0.4+ 0.1 k$, $\bgamma=  0\times \bone_q$;
(E4) $\sqrt{\theta}= 0 $, $\bgamma= (0.2k)\times \bgamma_{0} $;
(E5) $\sqrt{\theta} = 0.3+  0.3 k$, $\bgamma=   (-0.2k)\times \bone_q$;
(E6) $\sqrt{\theta}=   0.6k $,      $\bgamma=  (-0.45k)\times \bgamma_{0} $.
The prevalence  is still around 5\% in all scenarios.
\end{example}

Tables  \ref{case-ex-c1} and  \ref{case-ex-c2} reports the   simulation results.
We observe that the proposed  SS-MAX outperforms  MiST by a large margin
in scenarios (E2), (E5), and (E6)
although they have similar powers in scenarios (E1) and (E3).
In scenario (E4),  the MiST-Random cannot control its type I error when testing the existence of
the random effect,  therefore the large power of MiST may be misleading and risky.
Surprisingly, the three SS tests,
SS-MAX, SS$(\alpha_p)$ and SS$(\hat \alpha)$ test, have almost the same type I error and
powers when testing the random effect of rare variants.

\begin{table}[!http]
\caption{
Rejection probabilities (\%) of the  tests under consideration and the FS score test
for Scenarios (E1)-(E3) in Example \ref{ex-c}
($\bgamma_{0} =(\bone_{q/2}^\T,  \;  -\bone_{q/2}^\T)^\T$)
\label{case-ex-c1}}

\centering

\begin{tabular}{ l  rrrrrr  }\hline

  &  $k=0$&$k=1$ &$k=2$ &$k=3$ &   $k=4$&$k=5$    \\\hline
                         & \multicolumn{6}{c }{(E1) $\sqrt{\theta}= 0 $, $\bgamma=   (-0.08k)\times \bone_q $ }    \\

$\randomsc$--MAX         &  2.05&  2.35&  2.20&  2.90&  2.80&  3.65    \\
$\randomsc(\alpha_p)$    &  2.10&  2.35&  2.15&  2.15&  2.75&  2.95    \\
$\randomsc(\hat \alpha )$&  3.35&  3.35&  2.80&  3.60&  4.05&  3.60    \\
MiST-Random              &  4.30&  4.25&  3.95&  4.20&  3.95&  3.70    \\[1ex]

$\sumsc$--MAX            &  2.90&  7.10& 16.70& 35.10& 56.25& 76.95    \\
$\sumsc(\alpha_p)$       &  2.80&  8.10& 19.65& 39.10& 60.80& 79.75    \\
$\sumsc(\hat \alpha)$    &  3.15&  8.35& 20.70& 40.55& 62.10& 80.90    \\
Burden                   &  3.75& 10.65& 26.35& 48.50& 69.25& 87.25    \\
SKAT                     &  4.15&  6.55& 10.00& 20.85& 31.65& 52.05    \\
SKAT-O                   &  3.10&  9.80& 21.30& 41.60& 63.10& 81.65    \\
MiST                     &  3.55&  8.55& 18.20& 38.20& 57.50& 77.75    \\\hline

           &  \multicolumn{6}{c }{(E2) $\sqrt{\theta}= 0.8+ 0.3k $, $\bgamma=   (-0.25k)\times \bone_q $ }    \\

$\randomsc$--MAX         &     8.70& 17.70& 40.40&  58.10&  80.45&  92.50   \\
$\randomsc(\alpha_p)$    &    10.65& 21.85& 44.40&  58.75&  76.90&  89.30   \\
$\randomsc(\hat \alpha )$&    11.80& 18.25& 37.75&  62.45&  84.25&  93.60   \\
MiST-Random              &     5.40&  6.30&  4.95&   6.40&   7.05&   6.70   \\[1ex]

$\sumsc$--MAX            &    78.30& 70.40& 85.65&  94.50&  99.55& 100.00   \\
$\sumsc(\alpha_p)$       &    81.85& 74.50& 86.20&  94.15&  99.50&  99.95   \\
$\sumsc(\hat \alpha)$    &    82.30& 72.95& 85.15&  95.30&  99.55& 100.00   \\
Burden                   &    85.75& 73.25& 81.35&  89.50&  97.85&  99.60   \\
SKAT                     &    48.35& 37.70& 44.25&  59.05&  80.50&  93.10   \\
SKAT-O                   &    80.55& 67.45& 74.20&  85.30&  96.40&  99.35   \\
MiST                     &    76.55& 63.15& 69.50&  82.75&  95.05&  98.95   \\\hline

  &     \multicolumn{6}{c }{(E3) $\sqrt{\theta} = 0.4+   0.1k $, $\bgamma=  0\times \bone_q$} \\

$\randomsc$--MAX         &  2.05&  2.85&  3.70&  4.45&  8.70& 17.45   \\
$\randomsc(\alpha_p)$    &  2.10&  3.55&  4.15&  5.95& 10.65& 22.40   \\
$\randomsc(\hat \alpha )$&  3.35&  4.20&  5.60&  5.85& 11.80& 18.40   \\
MiST-Random              &  4.30&  4.80&  4.95&  4.30&  5.40&  5.05   \\[1ex]

$\sumsc$--MAX            &  2.90&  3.60&  9.35& 31.60& 78.30& 98.85   \\
$\sumsc(\alpha_p)$       &  2.80&  4.60& 10.10& 36.20& 81.85& 98.90   \\
$\sumsc(\hat \alpha)$    &  3.15&  4.90& 11.20& 36.00& 82.30& 98.95   \\
Burden                   &  3.75&  5.20& 11.90& 42.95& 85.75& 99.10   \\
SKAT                     &  4.15&  4.40&  5.90& 14.95& 48.35& 87.10   \\
SKAT-O                   &  3.10&  5.35& 10.50& 35.90& 80.55& 98.45   \\
MiST                     &  3.55&  4.90&  9.35& 32.30& 76.55& 98.05   \\\hline
\end{tabular}
\end{table}

\begin{table}[!http]
\caption{
Rejection probabilities (\%) of the  tests under consideration and the FS score test
for Scenarios (E4)-(E6) in Example \ref{ex-c}
($\bgamma_{0} =(\bone_{q/2}^\T,  \;  -\bone_{q/2}^\T)^\T$)
\label{case-ex-c2}}

\centering

\begin{tabular}{ l  rrrrrr }\hline
  &  $k=0$&$k=1$ &$k=2$ &$k=3$ &   $k=4$&$k=5$    \\\hline
   &      \multicolumn{6}{c }{(E4) $\sqrt{\theta}= 0 $, $\bgamma= ( 0.2k)\times \bgamma_{0} $} \\

$\randomsc$--MAX         &     2.05&  3.35&  2.65&   6.05&  12.85&  18.60      \\
$\randomsc(\alpha_p)$    &     2.10&  2.40&  2.95&   7.90&  12.25&  23.00      \\
$\randomsc(\hat \alpha )$&     3.35&  5.15&  4.45&   8.10&  16.85&  20.85      \\
MiST-Random              &     4.30& 32.15& 91.05& 100.00& 100.00& 100.00      \\[1ex]

$\sumsc$--MAX            &     2.90&  3.80&  3.85&  10.85&  36.50&  76.20      \\
$\sumsc(\alpha_p)$       &     2.80&  4.15&  4.25&  12.40&  38.50&  78.60      \\
$\sumsc(\hat \alpha)$    &     3.15&  5.05&  5.00&  13.10&  40.85&  77.90      \\
Burden                   &     3.75&  4.45&  4.55&  10.30&  28.20&  60.75      \\
SKAT                     &     4.15& 31.05& 89.55& 100.00& 100.00& 100.00      \\
SKAT-O                   &     3.10& 22.75& 84.45&  99.85& 100.00& 100.00      \\
MiST                     &     3.55& 23.50& 83.40&  99.85& 100.00& 100.00      \\\hline

 &     \multicolumn{6}{c }{(E5) $\sqrt{\theta} = 0.9+  0.3k $, $\bgamma=   (-0.2k)\times \bone_q$} \\

$\randomsc$--MAX         & 10.30& 14.70& 45.95& 60.75& 80.55&  93.05   \\
$\randomsc(\alpha_p)$    & 12.60& 18.40& 49.10& 59.65& 76.85&  89.55   \\
$\randomsc(\hat \alpha )$&  8.30& 14.65& 39.00& 62.65& 84.20&  94.75   \\
MiST-Random              &  5.45&  4.65&  6.00&  6.25&  8.00&   7.25  \\[1ex]

$\sumsc$--MAX            &  9.30& 11.85& 45.25& 77.95& 97.50&  99.95  \\
$\sumsc(\alpha_p)$       &  9.90& 12.75& 45.60& 76.25& 95.75&  99.90  \\
$\sumsc(\hat \alpha)$    &  8.00&  9.90& 37.80& 76.35& 98.00& 100.00  \\
Burden                   &  5.85&  4.20& 15.25& 53.25& 88.20&  99.25  \\
SKAT                     &  5.95&  4.40&  7.95& 24.20& 57.35&  90.40  \\
SKAT-O                   &  6.00&  4.45& 13.25& 47.10& 83.80&  98.90  \\
MiST                     &  5.90&  4.45& 12.35& 44.15& 80.50&  98.55  \\ \hline

 &      \multicolumn{6}{c }{(E6) $\sqrt{\theta}=  0.6k $,   $\bgamma=   (-0.45k)\times \bone_q$} \\

$\randomsc$--MAX         &   2.05&  4.20&  8.80& 18.80& 41.75& 62.50    \\
$\randomsc(\alpha_p)$    &   2.10&  5.05&  9.15& 21.50& 43.80& 63.95    \\
$\randomsc(\hat \alpha )$&   3.35&  3.85&  6.35& 17.40& 39.75& 61.75    \\
MiST-Random              &   4.30&  3.65&  5.20&  5.80&  5.15&  5.85   \\[1ex]

$\sumsc$--MAX            &   2.90& 42.00& 74.40& 67.70& 54.90& 53.20   \\
$\sumsc(\alpha_p)$       &   2.80& 45.75& 77.05& 69.75& 55.65& 51.80   \\
$\sumsc(\hat \alpha)$    &   3.15& 45.65& 76.00& 69.10& 52.30& 48.25   \\
Burden                   &   3.75& 52.85& 81.75& 69.45& 35.35&  6.20   \\
SKAT                     &   4.15& 22.55& 45.25& 34.50& 15.75&  6.10   \\
SKAT-O                   &   3.10& 46.30& 75.75& 63.20& 29.65&  6.55   \\
MiST                     &   3.55& 41.80& 72.05& 60.50& 28.00&  6.15   \\\hline
\end{tabular}
\end{table}

Overall the proposed  SS-MAX has similar performance to the ideal SS test,
in which the intercept $\alpha_p$ is assumed to be known,
and both have desirable control on type I error and
are more powerful than existing tests when \eqref{random-effect-model} is correct.
In addition,  the prevalence information can indeed
help to increase power.

\section{Application to GWAS of pancreatic cancer}

We demonstrated the performance of the proposed methods by
  applying it to two GWAS of pancreatic cancer.
The first GWAS (PanScan I) genotyped about 550000 SNPs from 1896
individuals with pancreatic cancer and 1939 controls drawn from
12 prospective cohorts and one hospital-based case-control study
\citep{Amundadottir2009}.
The second GWAS (PanScan II) genotyped about 620000 SNPs in 1679 cases and 1725 controls
 from seven case-control studies \citep{Petersen2010}.
For our analysis, we focused on people primarily of European ancestry, i.e.
people with their European admixture coefficient larger than 0.85 estimated by STRUCTURE
\citep{Pritchard2000}.
There were 3275 cases and 3376 controls left for our  analysis.

We conducted gene-based multiple locus analysis on the two GWAS data   combined.
We focused on genes in the PredictDB Data Repository that were defined by eQTL SNPs
identified by prediction models trained on the gene expression on pancreatic issues
from GTEx Version 7 data  \citep{Gamazon2015}.
There were a total of 4573 genes in the data repository.
We only considered eQTL SNPs identified in the prediction model for a given gene
because those SNPs had cis effect on the expression of the corresponding gene
and thus are more likely to be functional.

In the analysis, the logistic regression model was adjusted for study, age, sex and
the  10 principle components  (five from each of the two GWAS) for the adjustment of population stratification.
We focused on results from the following five tests,  SS--MAX, SS$(\hat\alpha)$,
SKAT, SKAT-O and  MiST. In  Table \ref{realdata}  we showed results on genes on which at least
one of the five considered tests generate a p-value less than $10^{-4}$.
We also highlighted p-values that are less than $1.1 \times 10^{-5}$,
which is the Bonferroni threshold to control the family wise error at
the level of 0.05 for a given test.
It appears from the table that each of the considered tests was able to identify one,
but not necessary the same, global significant gene, and p-values on a
given genes can be quite different. For example, the SS-MAX test produces
a highly significant p-value on gene IRX2, but p-values from other tests
on this gene are not significant at the Bonferroni threshold.

\begin{sidewaystable}
\caption{
Test  results of the top 6 genes, in which  at least one of
the SSMAX, SS$(\hat\alpha)$,  burden,  SKAT, SKAT-O and  MiST tests
produces a $p$-value less than $10^{-4}$
\label{realdata}}
\centering
\begin{tabular}{ l  cccccc   }   \hline
 Gene Name   &    IRX2  &   CAMK2N1&    QPCTL &  BCAR1   &   CTNNA2 &   ZDHHC11B   \\
Number of SNP&     107  &     32   &     18   &   20     &     9    &    13      \\  \hline
RS-MAX       & 0.0108053& 0.0030223& 0.5628025& 0.6292526& 0.0779794& 0.6150062  \\
RS$(\hat \alpha)$ & 0.5000000&\bf  0.0000101& 0.3515203& 0.5000000& 0.5000000& 0.3522865  \\
MiST-Random  & 0.0753358& 0.7621624& 0.3244760&\bf  0.0000051&\bf  0.0000052&\bf  0.0000018  \\[1ex]

SS-MAX       & \bf 0.0000070& 0.0033030& 0.0488375& 0.0494004& 0.1911246& 0.9909008  \\
SS$(\hat \alpha)$ & 0.0001058& {\bf 0.0000185} & 0.0354906& 0.0367571& 0.7623834& 0.8170036  \\
Burden       & 0.0745403& 0.3826072& 0.0000500& 0.0021725&\bf  0.0000087& 0.7191433  \\
SKAT         & 0.4618428& 0.6735554& 0.0000883& 0.0001503&\bf  0.0000087& 0.7346693  \\
SKAT-O       & 0.1271816& 0.5441588& 0.0000443& 0.0002321&\bf  0.0000087& 0.7352864  \\
MiST         & 0.0000364& 0.2925117& 0.0351248&\bf  0.0000015& 0.0000452& 0.0000249  \\\hline
\end{tabular}
\end{sidewaystable}

The proposed SS-MAX detects the gene IRX2, with p-value$=7\times 10^{-6}$,
which is much samller than  the p-value of MiST, 3.64$\times 10^{-5}$,
which also provides evidence for the significance of gene IRX2.
The SS$(\hat \alpha)$  identifies the gene CAMK2N1 with a p-value $1.85\times 10^{-5}$,
which is mainly contributed by RS$(\hat \alpha)$ with a p-value $1.01\times 10^{-5}$.
The significant or nearly significant p-values of MiST on
genes BCAR1, CTNNA2, and ZDHHC11B are probably
due to the signal detected by MiST-Random.
However, according to our simulation experience,
MiST-Random tends to have inflated type I errors
when the assumed model in \eqref{random-effect-model} is violated.
Burden, SKAT, and SKAT-0 all detect the gene CTNNA2.

\begin{figure}
\begin{center}
\makebox{
\includegraphics[width=0.3\textwidth]{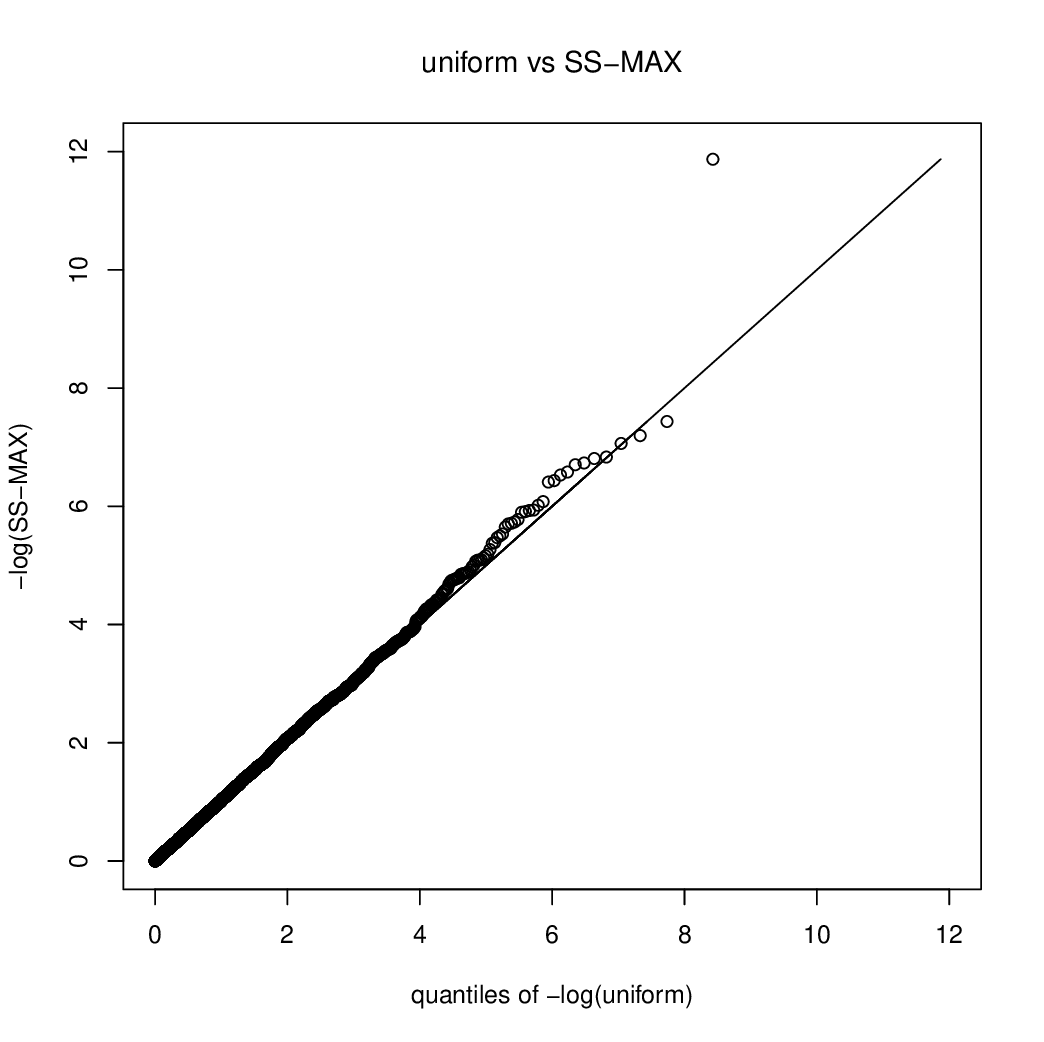}
\includegraphics[width=0.3\textwidth]{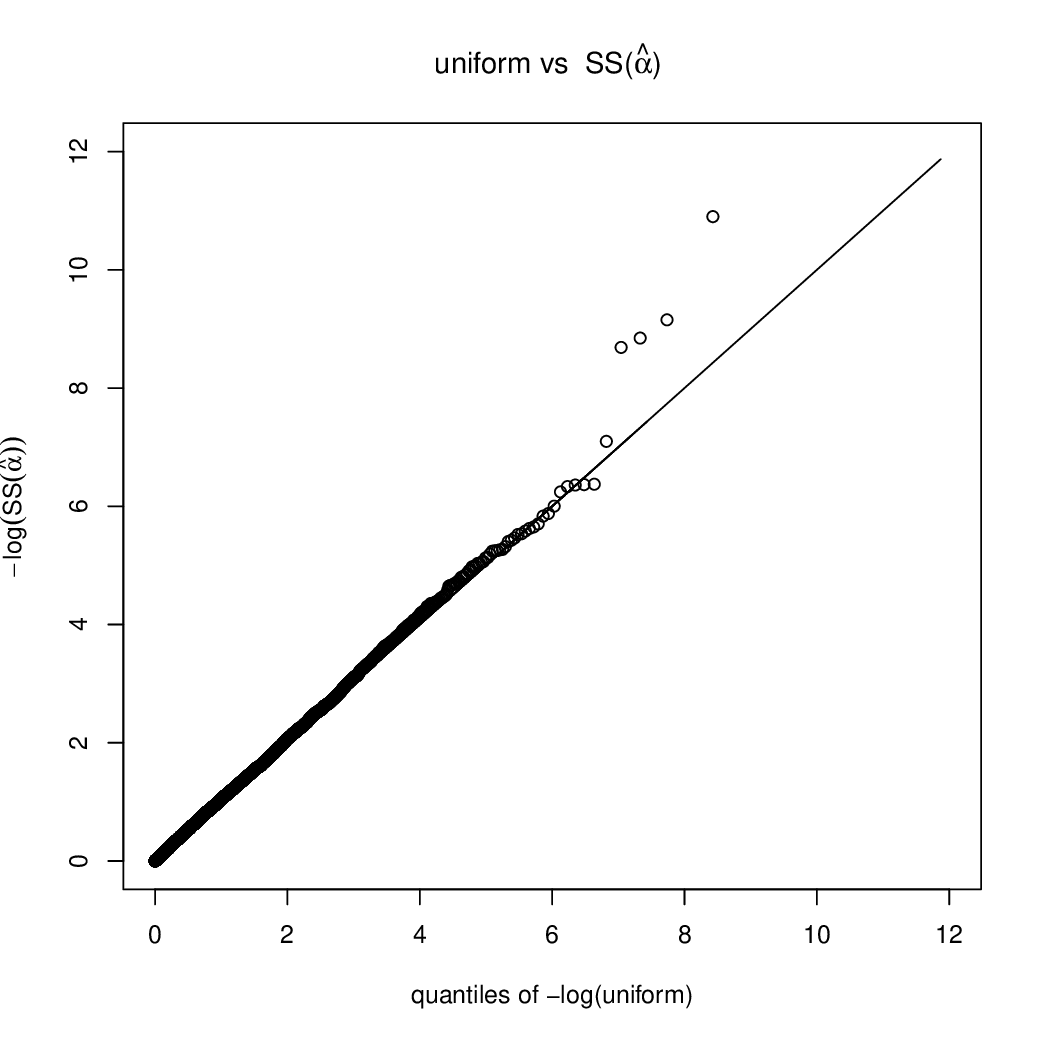}
\includegraphics[width=0.3\textwidth]{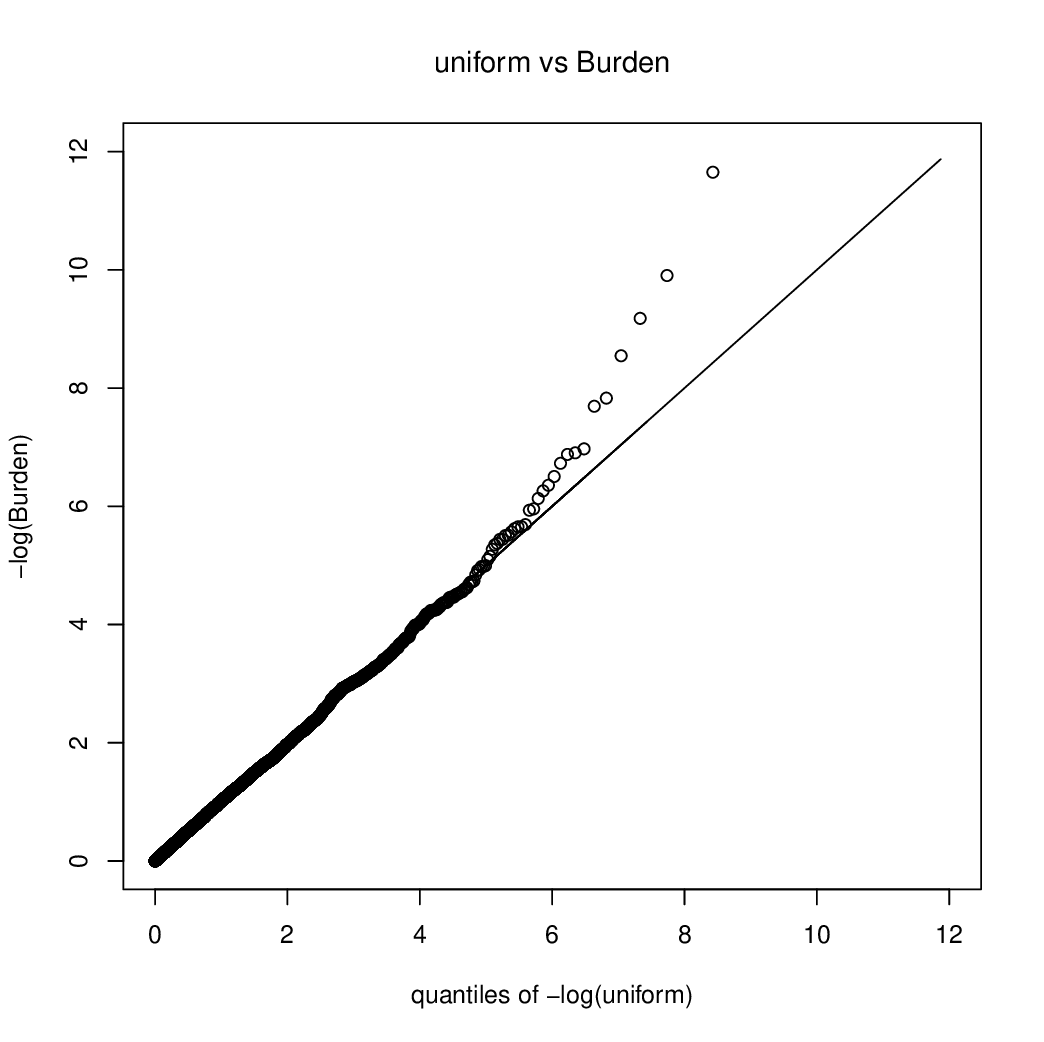}
}
\\
\makebox{
\includegraphics[width=0.3\textwidth]{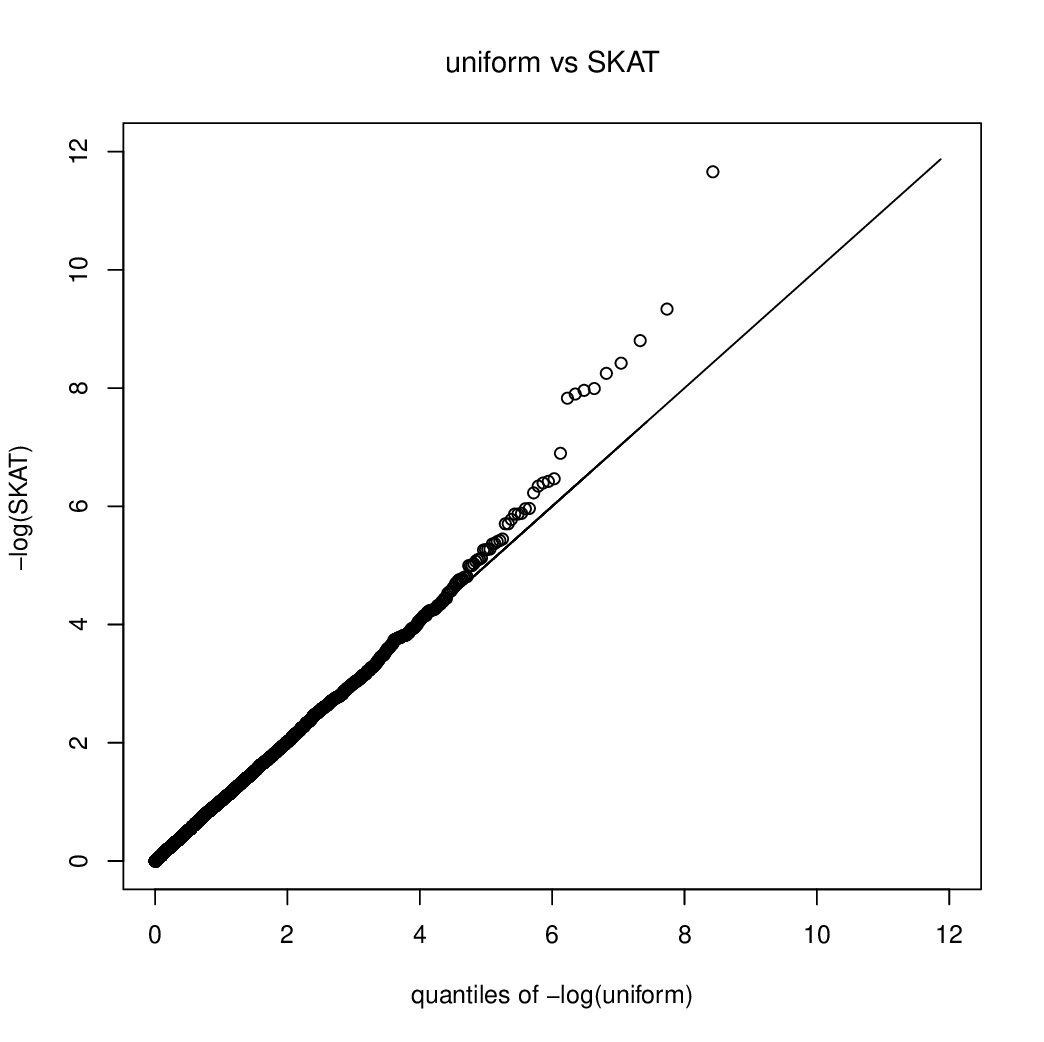}
\includegraphics[width=0.3\textwidth]{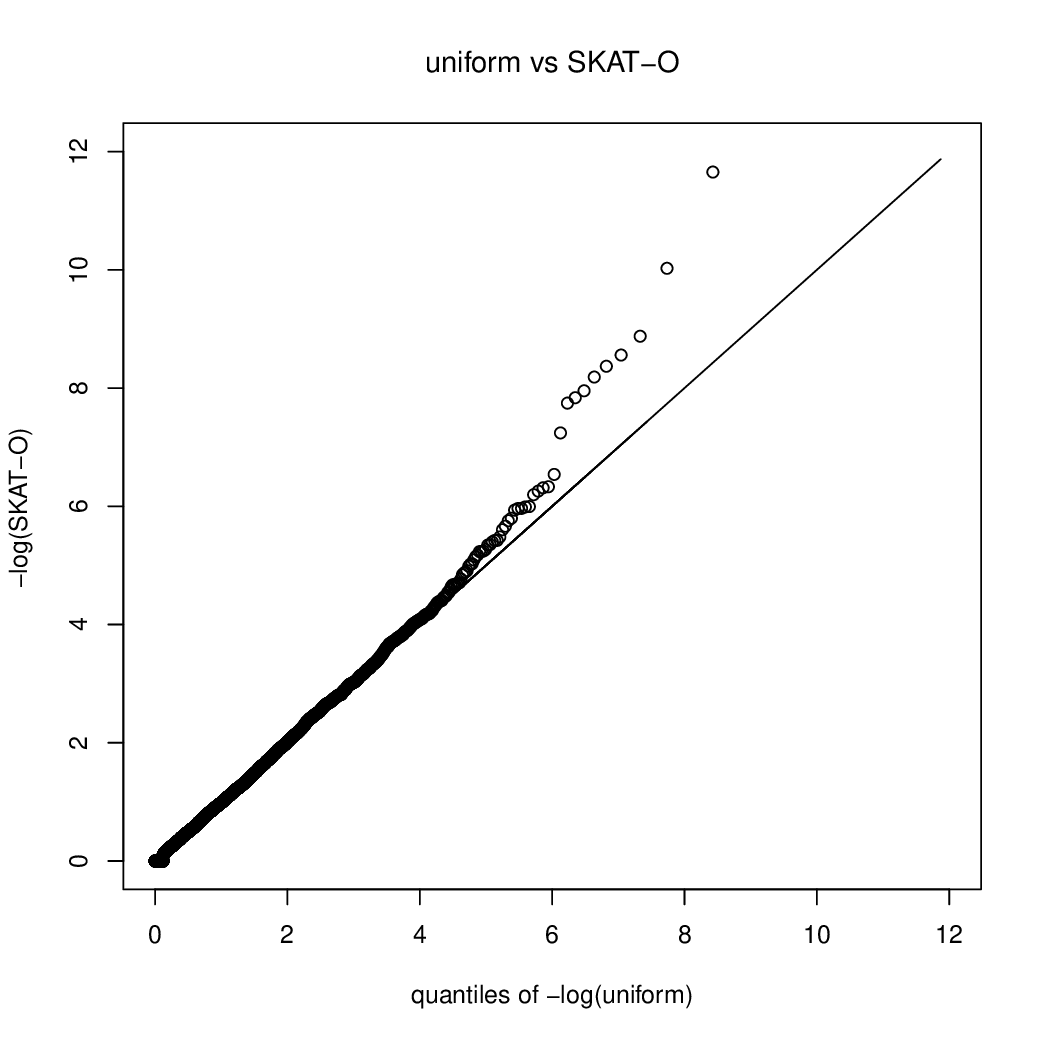}
\includegraphics[width=0.3\textwidth]{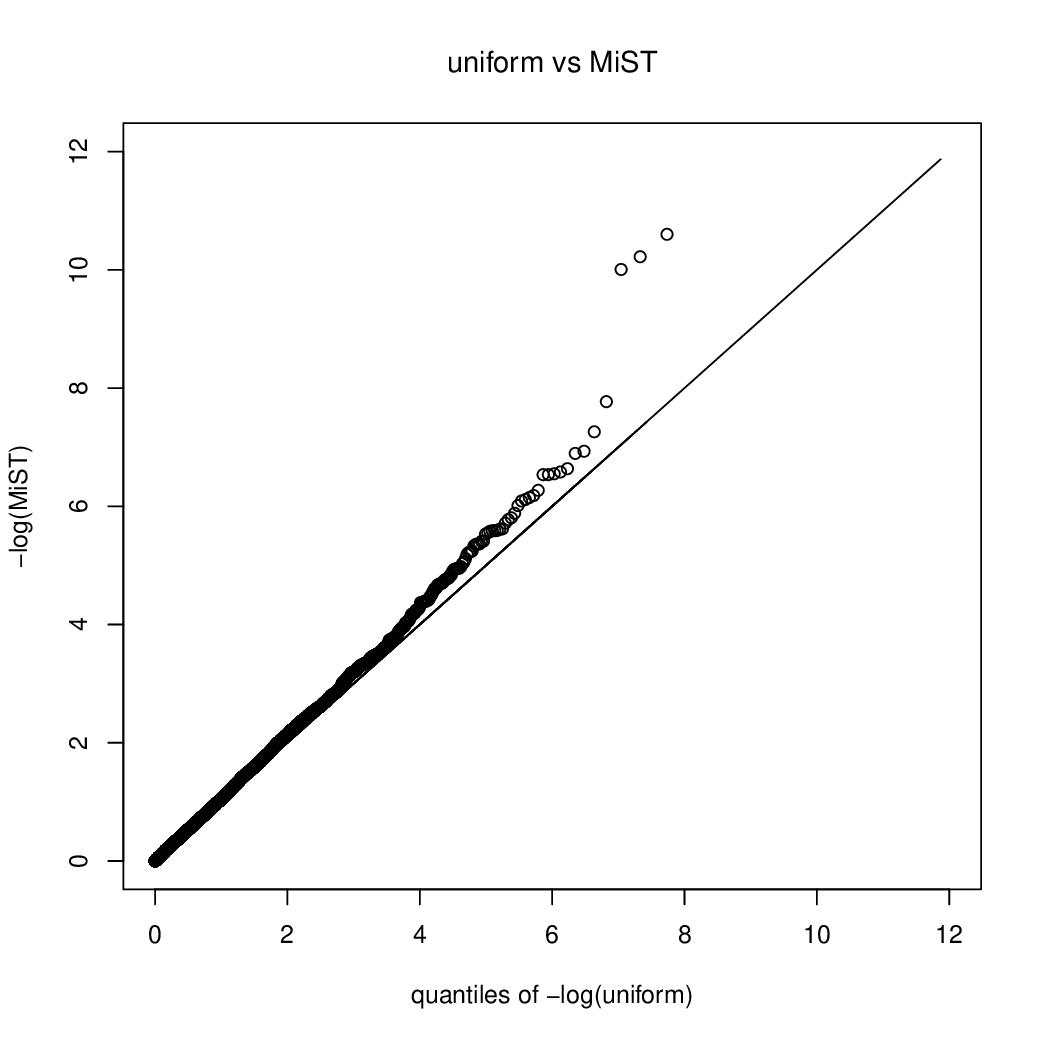}
}
\end{center}
\caption{qq-plots of minus logarithm of p-values. \label{fig-qqplot}}
\end{figure}

We also display in Figure \ref{fig-qqplot} qq-plots of all considered tests.
Under the assumption that most of the genes are not associated with the outcome,
we would expect the qq-plot based on p-values for 4573 considered genes under
each test should align well with the diagonal line.
From Figure \ref{fig-qqplot},
we see the two proposed tests SS-MAX and SS$(\hat \alpha)$ have the expected patterns
in their qq-plots, however the burden, SKAT, SKAT-O, and MiST tests
fail more or less.
In addition,  points at the upright corner of each qq-plot
correspond to  the genes that are most likely significant.
For example,  the outlier point at the upright corner of the qq-plot of SS-MAX
correspond to the gene IRX2 reported in Table  \ref{realdata}.

\section{Discussion}
\label{s:discuss}

In this paper we have systematically derived the score based tests based on prospectively
and retrospectively collected data and studied their large-sample behaviors.
We have made clear explicitly the similarity and difference
between the two different likelihoods.
Our general finding is that the sampling scheme should be taken into account
in likelihood-based inference.
Specifically, we have found the retrospective likelihood
based on case and control data can produce more powerful score tests
than blindly using of prospective likelihood in genetic association
studies with random effects.
In contrast to the well-known fact that disease prevalence does not
carry any information for the odds ratio parameter estimation
in the case and control study,
we can use this information to enhance the test power;
sometimes, the gain is overwhelming.
Instead of the common random-effect model used in   \cite{Lin1997}
in which all individuals share a common random effect,
we have used a conventionally used random effect model
where observations are independent unless they come from the same individual.
To test the fixed and random effect together,
we have proposed an innovative combination method by taking into consideration
the fact that score test for the random effect is one sided
and that for the fixed effect is two sided.
Different methods to calculate the p-values are also discussed.
Simulation studies show overall advantages of the newly proposed tests
over the existing ones.
Some new findings are discovered from the analysis of two GWAS pancreatic cancer data sets.
Clearly our methodology can be generalized to
other case and control data problems as well as
the choice based sampling econometric problems.

\end{document}